\documentclass[superscriptaddress,nofootinbib,a4paper,11pt]{article}
\usepackage{jheppub}
\usepackage{graphicx}
\usepackage{subcaption}
\usepackage{float}
\usepackage{amsmath}
\usepackage{amssymb}
\usepackage[utf8]{inputenc}
\usepackage{hyperref}
\usepackage{color}
\usepackage{slashed}
\usepackage{verbatim}
\usepackage{multirow} 
\usepackage{tikz}
\usepackage{tikz-feynman}
\usepackage{picture}
\usepackage{pst-plot}
\usepackage{graphics}

\allowdisplaybreaks

\newcommand{\be}{\begin{equation}}
\newcommand{\ee}{\end{equation}}

\begin{document}

\title{\boldmath Can Effective $4-$Quark Operators Describe Signals of
  a Supersymmetric Diquark Model at the LHC?}

\author[a]{Manuel Drees,}
\author[a,*]{Cong Zhang\note[*]{Corresponding author.}}

\affiliation[a]{Bethe Center for Theoretical Physics and Physikalisches
  Institut, Universit\"at Bonn,\\Nussallee~12, D-53115 Bonn,
  Germany}

\emailAdd{drees@th.physik.uni-bonn.de}
\emailAdd{zhangcong.phy@gmail.com}

\abstract{The Standard Model Effective Field Theory (SMEFT) is
  constrained by current LHC data. Supposedly extensions of the
  Standard Model (SM) involving heavy particles can be constrained by
  matching onto the SMEFT. However, the reliability of these indirect
  constraints compared to those derived directly from the UV model
  remains an open question. In this paper, we investigate whether
  $4-$quark operators can accurately capture the effects of an
  $R-$parity-violating (RPV) supersymmetric model on the production of
  pairs of top quarks, for parameters that satisfy all known
  constraints and lead to measurable effects. We assume that the
  sbottom is the lightest supersymmetric particle and focus on its
  interaction with a light quark and a top quark; the sbottom thus
  acts like a specific diquark. The $4-$quark operators arise by
  integrating out the sbottom at tree level. We analyze measurements
  of inclusive top pair production by the CMS and ATLAS
  collaborations. We find that the $4-$quark operators can accurately
  describe the RPV model's effects only for very heavy sbottom
  squarks, where the effects are well below the sensitivity of LHC
  experiments for all values of the RPV coupling that satisfy
  unitarity constraints. Therefore present or near-future bounds on
  this RPV model can {\em not} be derived from SMEFT analyses.}


\maketitle

\section{Introduction}
\label{sec:intro} 

More than 15 years after the LHC experiments started to take data they
have not discovered a single particle not described by the Standard
Model (SM). This is often seen as an argument against extensions of
the SM that were designed to address the electroweak hierarchy problem
by introducing new particles at the weak scale, e.g. by postulating
that nature is supersymmetric (see e.g. \cite{Drees:2004jm,
  Baer:2006rs}) or that additional spatial dimensions exist
\cite{Antoniadis:1998ig, Randall:1999ee}.

Instead, a supposedly model--independent, or at least less
model--dependent, approach has become popular in the last decade or
so. Here one assumes that no new particles exist below the TeV scale.
At energies well below a TeV effects of new particles can then be
described by an effective theory (EFT), which extends the SM by a set
of non--renormalizable higher--dimensional operators. EFTs of this
kind had previously proven very helpful. For example, weak decays of
charm-- or beauty--flavored mesons and baryons can be described by a
low--energy EFT which respects the $SU(3)_C \times U(1)_{\rm em}$
symmetry of the SM but describes $W$ and $Z$ exchange, as well as
loops involving top quarks, through a set of higher--dimensional
operators. Among other things, this simplifies the computation
of QCD corrections (see e.g. \cite{Buchalla:1995vs}). Similarly, it
is clear that baryon-- and lepton--number violating interactions
due to the exchange of gauge bosons predicted by Grand Unified
theories (see e.g. \cite{Ross:1985ai}) can safely be described by
an EFT at experimentally accessible energies \cite{Weinberg:1979sa}.

These considerations led to the development of the Standard Model
Effective Field Theory (SMEFT) \cite{Buchmuller:1985jz,
  Grzadkowski:2010es, Brivio:2017vri} as a framework for
systematically probing potential new physics beyond the Standard Model
(BSM). It assumes that all interactions respect the full SM gauge
symmetry, based on the group $SU(3)_c\times SU(2)_L\times U(1)_Y$.
Since the SM already contains a sizable number of particles, well over
$2,000$ new terms, with independent coefficients, can be constructed
already at energy dimension $d=6$ if no further simplifying
assumptions are made.

Phenomenological investigations of the SMEFT therefore have focused
on relatively small subsets of these new operators. Examples are
explorations of the top quark sector \cite{Buckley:2016cfg,
  Buckley:2015lku, Hartland:2019bjb, Brivio:2019ius, Ellis:2020unq,
  Bissmann:2020mfi}, Higgs and electroweak precision data
\cite{Biekotter:2018ohn, Ellis:2018gqa, daSilvaAlmeida:2018iqo}, gauge
boson production \cite{Baglio:2020oqu, Alioli:2018ljm, Ethier:2021ydt,
  Greljo:2017vvb}, vector boson scattering \cite{Ethier:2021ydt,
  Gomez-Ambrosio:2018pnl, Dedes:2020xmo}, as well as various
low--energy constraints \cite{Aebischer:2018iyb, Falkowski:2019xoe,
  Falkowski:2017pss, Bruggisser:2021duo}. In the analysis of the top
quark sector, the Minimal Flavor Violation (MFV) hypothesis
\cite{DAmbrosio:2002vsn} is usually adopted as the baseline
scenario.

A rather ambitious more recent study \cite{Ethier:2021bye} provides a
global interpretation of Higgs, diboson, and top quark measurements
from the LHC based on $50$ $d=6$ operators. By combining all input
data, the individual and global $95\%$ confidence level intervals are
obtained for all $50$ coefficients, using either linear or linear plus
quadratic SMEFT calculations. ``Linear'' here means that only terms
linear in the new Wilson coefficients are considered in the squared
matrix element for any given process; such contributions occur if a
SMEFT contribution interferes with an SM contribution. In contrast, a
``linear plus quadratic'' fit also includes terms bilinear or
quadratic in the new Wilson coefficients. This gives sensitivity to
contributions that do not interfere with the SM, e.g. to new color
structures. However, these ``quadratic'' terms are
${\cal O}(\Lambda^{-4})$, i.e. they show the same dependence on the
``new physics'' energy scale $\Lambda$ as contributions that are
linear in the Wilson coefficients of $d=8$ operators. Since the latter
are not considered in the fit, a quadratic $d=6$ fit is not consistent
from a power counting point of view.

One of the arguments in favor of SMEFT fits is that they should allow
to directly read off constraints on the parameters of renormalizable
models that predict new, heavy particles. To this end one only needs
to match the model to the SMEFT, in order to establish the relations
between the masses of couplings of the new particles proposed in a
given model and the relevant SMEFT coefficients. Bounds on the latter
then directly translate into bounds on the model parameters. This
sounds straightforward; however, in practice this procedure may fail
for a variety of reasons:

\begin{itemize}

\item Tree--level matching to the SMEFT basically amounts to shrinking
  propagators of new, heavy particles to a point. This can only work
  if the absolute value of the squared momentum flowing through this
  propagator is much smaller than the squared mass of the exchanged
  particle. This needs to be true in {\em all} events considered. At
  $e^+e^-$ colliders this will be true if the total Mandelstam$-s$ is
  much smaller than the squared mass $M^2$ of the new exchange
  particle, i.e. this criterion should be satisfied as long as
  $\sqrt{s} \leq 3 M$.  Of course, this statement also holds for $pp$
  or $p \bar p$ colliders, if $\sqrt{s}$ is the hadronic
  center--of--mass (cms) energy. However, when writing the SMEFT
  Wilson coefficients as $1/\Lambda_i^2$, fits of current LHC data
  typically lead to bounds on the $\Lambda_i$ not much above $1$ TeV;
  energies of this order of magnitude can easily be reached even in
  the partonic cms. It is therefore not clear a priori whether the
  SMEFT approximation is indeed applicable for values of the
  $\Lambda_i$ near present or even future LHC sensitivity.\footnote{In
    principle one can try to ensure applicability by imposing
    kinematic cuts that limit the momentum flow through the relevant
    massive propagators. However, this will remove those events which
    are {\em most} sensitive to the existence of heavy new
    particles. This procedure will therefore certainly not give the
    real bounds on any concrete model that could be derived from LHC
    data.}
  
\item A concrete renormalizable model will in general only generate a
  subset of SMEFT operators, at least at tree level. Moreover, it may
  impose relations between the SMEFT Wilson coefficients. This means
  that a global SMEFT fit, which allows (many) more operators than are
  actually generated by a given model, will usually lead to (much)
  weaker constraints on the coefficients of the operators that
  actually are generated, since the many parameter SMEFT fit allows
  for cancellations that may not be possible in any given model. This
  is true even if one finds a SMEFT fit that only considers the
  operators that are generated in the model of interest, unless the
  SMEFT fit also imposes the relations between the coefficients of
  these operators that follow for the given model. Since there's in
  principle an uncountable infinity of such relations. This
  combinatorial problem means that in practice one will (almost) never
  find a SMEFT fit that actually has the right number of degrees of
  freedom to describe a given UV complete model. This difficulty arises
  also in analyses of data from $e^+e^-$ colliders, where annihilation
  events have an (almost) fixed center--of--mass energy (barring events
  with hard initial state radiation).

\item A model may not generate {\em any} $d=6$ SMEFT operators at tree
  level; examples include supersymmetric models with conserved
  $R-$parity, and extra dimensional models with conserved KK parity.
  The actual LHC constraints on such models nearly always come from
  searches for the production of pairs of new particles; these
  constraints cannot be captured by SMEFT analyses.\footnote{Some of
    recent studies have explored the validity of SMEFT at the LHC
    analyzing loop--induced effects, such as a dark matter model
    with a $\mathcal{Z}_2$ symmetry \cite{Lessa:2023tqc}, as well as
    non--degenerate stop squarks within the context of Higgs couplings
    \cite{Drozd:2015kva}.}

\end{itemize} 

Very similar problems were encountered when, several years before the
SMEFT, a ``WIMP effective theory'' was suggested \cite{Bai:2010hh,
  Beltran:2010ww} aiming for a ``model-–independent'' description of
monojet (and, more generally, ``mono$-X$'') searches at the Tevatron
and LHC. However, it became clear after a while that most UV--complete
theories describing WIMP production at hadron colliders cannot be
described by an effective field theory in the experimentally
accessible parameter space. Instead, the real bounds typically come
from searches for the on--shell production of the relevant
mediator(s); see e.g.  \cite{Fox:2011pm, Goodman:2011jq,
  Belwal:2017nkw, Drees:2019iex}

In this paper, we study the validity of the SMEFT description of an
$R-$parity violating (RPV) supersymmetric model \cite{Barbier:2004ez}.
Specifically, we analyze measurements of inclusive top pair events,
where the new particle contributes at tree level.

The RPV model is a simple extension of the minimal supersymmetrized
standard model (MSSM) by adding extra superpotential terms that break
baryon or lepton number (but not both, since that would lead to rapid
proton decay). Searches for superparticles in the framework of this
model have been carried out by both the ATLAS and CMS collaborations
\cite{ATLAS:2021fbt, ATLAS:2020wgq, ATLAS:2019fag, ATLAS:2018umm,
  CMS:2021knz, CMS:2018skt, CMS:2017szl, CMS:2016vuw, CMS:2016zgb,
  CMS:2013pkf, ATLAS:2015gky, ATLAS:2015rul}. Here we consider a
scenario where the superpartner of the right--handed $b$ quark, dubbed
$\tilde b_R$, couples to a top quark and a light quark through a
baryon number violating interaction. For simplicity we assume that all
other superpartners are considerably heavier than $\tilde b_R$, so
that their effect on LHC physics is negligible. Our model can thus be
considered to be a particular example of a diquark model. By
integrating out $\tilde b_R$ some $4-$quark operators are generated;
of course, these operators are part of the SMEFT set.

Our main goal is to find out whether the SMEFT description in terms of
these $4-$quark operators can provide reliable constraints on the RPV
model through the analysis of inclusive top pair events. As mentioned
above, a SMEFT description should indeed become possible for
sufficiently heavy $\tilde b_R$. However, it is a priori not clear
whether this is true for $\tilde b_R$ masses and couplings near
present or near--future sensitivity; note that the relevant coupling
is bounded from above by independent (theoretical) arguments.

In this RPV model new contributions to inclusive top pair production
primarily arise from two sources: top pair production via sbottom
exchange in the $t-$channel, as well as top pair plus single jet
production from diagrams with a potentially on--shell sbottom in the
intermediate state. We will see that in the latter case the
$\tilde b_R$ tends to be produced on--shell even for masses up to $3$
TeV (with coupling around 1); this contribution cannot be described by
the SMEFT. Nevertheless the SMEFT might still work for inclusive
$t \bar t$ production, if this second contribution is small. Conversely,
even if this contribution is small, some minimal sbottom mass is
required for the first contribution to be described accurately by
the SMEFT; only in this case will the limits derived within SMEFT
be applicable to the RPV model.

The remainder of this paper is structured as follows. In
Sec.~\ref{sec2} we review the theoretical framework of the RPV model
and investigate its tree--level matching to $4-$quark operators. In
Sec.~\ref{sec4} we briefly discuss the search for the production of
on--shell $\tilde b_R$ squarks. In Sec.~\ref{sec3} we perform a
detailed comparison between the RPV model and $4-$quark operators in
terms of differential distributions and the resulting constraints
derived from the measurements of inclusive top pair events from CMS
and ATLAS.  Finally we present our conclusions in Sec.~\ref{sec5}.

\section{Theoretical framework}
\label{sec2}

If $R-$parity is not imposed the superpotential of the minimal
supersymmetric standard model (MSSM) can contain the baryon number
($B$) violating terms \cite{Barbier:2004ez},
\begin{equation} \label{W_RPV}
  W_{\not R_p} = \frac{1}{2} \sum_{i,j,k} \lambda_{i j k}^{\prime \prime}
  U_i^c D_j^c D_k^c\,,
\end{equation}
where $i$,$j$,$k$ are generation indices. $SU(3)$ gauge invariance enforces
antisymmetry of the coupling,
$ \lambda^{\prime\prime}_{i j k}=- \lambda^{\prime\prime}_{i k
  j}$. The corresponding piece of the Lagrangian is
\begin{equation}\label{action}
  \mathcal{L}_{U_i^c D_j^c D_k^c} = - \frac{1}{2} \sum_{i,j,k}
  \lambda^{\prime \prime}_{i j k}\left( \tilde{u}_{i R}^\star \bar{d}_{j R}
    d_{k L}^c + \tilde{d}_{k R}^\star \bar{u}_{i R} d_{j L}^c
    + \tilde{d}_{j R}^\star \bar{u}_{i R} d_{k L}^c\right)+\text { h.c. }\, .
\end{equation}

We are interested in inclusive top pair production at the LHC. The
Lagrangian (\ref{action}) can contribute to this via terms coupling a
light quark, a top quark and some right--handed down--type squark.  In
order to maximize this contribution we want the light quark to be a
$d_R$ (rather than $s_R$, which has considerably smaller parton
density in the proton). In order to minimize $R-$parity conserving
production of the intermediate squark via the exchange of a gaugino in
the $t-$ or $u-$channel we select $\tilde b_R$ as exchanged
squark. The relevant term in eq.(\ref{action}) is thus the one
proportional to $\lambda^{\prime\prime}_{313}$. Moreover, we assume
that all other superparticles are considerably heavier than the
right--handed sbottom, and hence play no role for LHC physics.  It
should be mentioned that assuming a single scalar superparticle to be
much lighter than all the others isn't very natural from the model
building point of view; for example, larger gaugino masses can easily
turn the squared mass of the scalar particle negative at only slightly
larger energy scales. However, introducing additional relatively light
particles would make it less likely that the scenario can be described
by the SMEFT.

In order to keep things simple we also assume that
$\lambda^{\prime\prime}_{313}$ is the only sizable new
coupling.\footnote{This scenario has been explored in
  ref.~\cite{Allanach:2013qna} for $7-$TeV LHC data.} This also
relaxes some constraints. For example, the measurements of flavor
changing neutral currents (FCNC), e.g. in meson mixing, generally
impose constraints on the products of two different RPV couplings
\cite{Barbier:2004ez}. Furthermore, neutron--antineutron oscillations
are suppressed if gauginos are very heavy \cite{Barbier:2004ez,
  calibbi2017baryonnumberviolationsupersymmetry}. The remaining
constraint on $\lambda^{\prime\prime}_{313}$ arises from perturbative
unitarity, which requires that the coupling remains perturbative up to
some large energy scale $M_X$,
 \begin{equation}\label{unitary}
 \frac{ \left( \lambda^{\prime\prime}_{313}\right)^2(M_X)}{(4\pi)^2} < 1\,.
\end{equation}
In our case, setting $M_X \simeq 2 \cdot 10^{16}$ GeV (the scale of
supersymmetric Grand Unification), this requirement leads to
\cite{Barbier:2004ez}
\begin{equation} \label{uni-bound}
  \lambda^{\prime\prime}_{313}(1 \ {\rm TeV}) <1.12\,,
\end{equation}
almost independently of the other parameters of the theory. This
constraint implies that our RPV coupling cannot be larger than the
$SU(3)$ gauge coupling.

Since this is quite a strong restriction, we also consider the weaker
constraint from the requirement that the $\tilde b_R$ decay width does
not become too large, which would jeopardize the validity of
perturbation theory at the LHC energy scale. In our scenario the
sbottom decays only into a top antiquark and a down antiquark, with
decay width
\begin{equation}\label{decay}
  \Gamma_{\tilde b} =\frac { \left(\lambda^{\prime\prime}_{313} \right)^2
    \left( M^2_{\tilde b} - m_t^2 \right)^2 } {8\pi M_{\tilde b}^3}\,.
\end{equation}
For $\lambda^{\prime\prime}_{313}=3.0$,
$\Gamma_{\tilde b}/M_{\tilde b}$ approaches $0.35$ for large sbottom
mass, which we consider the upper limit of what is acceptable for a
``particle''. For $\lambda^{\prime\prime}_{313}=4.0$, the ratio
approaches $0.63$, i.e. the width exceeds half the mass. Ignoring
$\Gamma_{\tilde b}$ when integrating out the sbottom, as one typically
does when deriving the SMEFT limit of the theory, is then quite a poor
approximation. We will therefore always require
$\lambda^{\prime\prime}_{313}\leq 4.0$ when computing cross sections,
and often impose the stronger bound (\ref{uni-bound}).

Clearly $\tilde b_R$ exchange mediates interactions between
right--handed down and top quarks. The SMEFT limit is obtained by
integrating out the heavy sbottom. At tree level only two
dimension$-6$ operators in the Warsaw basis are generated:
\begin{equation}\label{wilson}
\begin{aligned}
  & \mathcal{O}_{t d}^{(1)} = \left( \bar{t} \gamma^\mu t\right) \left(
    \bar{d} \gamma_\mu d \right)\,; \\
  & \mathcal{O}_{t d}^{(8)} = \left( \bar{t} \gamma^\mu T^A t\right) \left(
    \bar{d} \gamma_\mu T^A d\right)\,,
\end{aligned}
\end{equation}
where $d$ and $t$ refer to right--handed down and top quarks in the
notation of ref.~\cite{Grzadkowski:2010es} and the $T^A$ are the
generators of $SU(3)$ in the fundamental representation. The
corresponding Wilson coefficients are
\begin{equation}\label{wilson1}
\begin{aligned}
  &C_{t d}^1 \equiv  \left( C_{u d}^{(1)}\right)_{3311} = \frac{
    \left| \lambda_{313}^{\prime \prime}\right|^2 } {3 M_{\tilde b}^2}\,; \\
  &  C_{t d}^8 \equiv \left( C_{u d}^{(8)} \right)_{3311} =
  - \frac{ \left| \lambda_{313}^{\prime \prime}\right|^2} {M_{\tilde b}^2}\,.
\end{aligned}
\end{equation}
As mentioned above, we ignored the sbottom decay width when
integrating out $\tilde b_R$. In order to capture the physics of our
RPV model using the SMEFT, only the $4-$quark operators in
eq.(\ref{wilson}) should be considered; eq.(\ref{wilson1}) shows that
their Wilson coefficients should satisfy
$C^1_{td} = - C^8_{td}/3 > 0$. The Wilson coefficients of all other
operators should be set to zero.

This illustrates the second possible problem with using the SMEFT as
stand--in for concrete models mentioned in the Introduction: it is
exceedingly unlikely that someone will have performed a SMEFT fit for
us that obeys all the necessary relations; indeed, we are not aware of
any such fit in our case. However, it should be noted that the
operator $\mathcal{O}_{t d}^{(1)}$ does not interfere with the leading
order QCD contribution, $d \bar d \rightarrow t \bar t$, which
proceeds via gluon exchange in the $s-$channel and thus requires the
initial and final quark bilinears to be in color octet states. A
``linear'' SMEFT fit (in the notation of the Introduction) that
ignores electroweak interactions therefore only needs to consider the
operator ${\cal O}^{(8)}_{td}$ when considering the impact on the
production of top pairs.

Of course, such a SMEFT description can only work if the SMEFT gives
a good approximation for the relevant kinematic distributions; before
checking whether this is the case for the differential top pair production
cross sections measured at the LHC, we briefly discuss possible bounds on
our model from direct searches for new particles.

\section{Direct search}
\label{sec4}

The search for pair production of new heavy particles, each
decaying to a top and a light quark (or gluon) jet, could lead to a
lower limit on the sbottom mass. Such a search has been performed by
the CMS collaboration, based on data from $13$~TeV collisions
corresponding to and integrated luminosity of 138 $\text{fb}^{-1}$
\cite{CMS:2024zpb}. They interpreted this as search for the pair
production of spin$-1/2$ excited top quarks $t^*$ with
$t^* \rightarrow t + g$ decay. This study provides the most stringent
limits on excited top quarks to date, superseding previous
measurements \cite{CMS:2017ixp, CMS:2013tlq}. However, since a deep
neural network was used to define the signal, we can't directly recast
their analysis for the sbottom case; not only is the overall cross
section (for fixed mass) considerably smaller for scalar (rather than
fermionic) color triplets, the angular distributions of the heavy
particles differ in the two cases.
 
Nevertheless a simple comparison between the observed limits for the
excited spin$-1/2$ top quark presented in Ref.~\cite{CMS:2024zpb} and
the predicted total cross section of sbottom pair production indicates
that this search might not be very sensitive to our RPV model. For
example, the observed limit on
$\sigma(p p \rightarrow t^*\bar{t^*}) \cdot
\mathcal{B}^2(t^*\rightarrow t g)$ is at $0.12$ pb for $M_{t^*} = 700$
GeV. In our simulation the pair production cross section at
$M_{\tilde b} = 700$ GeV is only $0.041$ pb, well below the observed
limit. The observed limit falls to $0.8$ fb for $M_{t^*} = 3$ TeV, but
the predicted $\tilde b_R \tilde b_R^*$ production cross section
decreases even more rapidly with increasing mass. Data taken at
$\sqrt{s} = 13$ TeV are therefore probably only sensitive to
$\tilde b_R$ pair production in our RPV model for
$M_{\tilde b} \lesssim 500$ to $600$ GeV. In the following we therefore
consider $M_{\tilde b} \geq 500$ GeV in our analyses.
 
\section{Validity of $4-$quark operator description of inclusive
  top pair events }
\label{sec3}

\subsection{Inclusive top-pair events}
\label{sec3.1}

Several analyses have used data on inclusive $t\bar{t}$ production in
order to constrain some SMEFT Wilson coefficients
\cite{Buckley:2015lku, Hartland:2019bjb, Brivio:2019ius,
  Ellis:2020unq, Ethier:2021bye}. To that end, typically published
parton--level or particle--level cross sections differential in
various kinematic variables were used, such as the invariant mass of
the top pair, the transverse momentum of a top quark or of the pair,
and the charge asymmetry of the top pair system. We use the same
kinematic distributions for a detailed comparison between the
predictions of the RPV model and those of the SMEFT, which uses
$4-$quark operators to describe the new contribution. We also compare
the limits that can be derived in these two frameworks.

We perform these comparisons for two LHC data sets covering $pp$
collisions at $\sqrt{s}=13$ TeV. The first set comes from measurements
of various differential cross sections of inclusive $t\bar{t}$
production by the CMS collaboration \cite{CMS:2021vhb}, based on an
integrated luminosity of $137$ fb$^{-1}$. This includes parton level
distributions over the full kinematic range, which are well suited for
our purpose. The second data set concerns the $t \bar t$ charge
asymmetry measured by the ATLAS collaboration with an integrated
luminosity of $139$ fb$^{-1}$ \cite{ATLAS:2022waa}.

\begin{figure}[htbp]
  \centering
    \begin{subfigure}{0.13\linewidth}
    \centering
\begin{tikzpicture}
    \begin{feynman}
        \vertex (i1) at (-1.7, 1.25) {$\bar{d}$};
        \vertex (i2) at (-1.7, -1.25) {${d}$};
        \vertex (f1) at (1.7, 1.25) {$t$};
        \vertex (f2) at (1.7, -1.25) {$\bar{t}$};
        \vertex (v1) at (0, 1.25);
        \vertex (v2) at (0, -1.25);
        
        \diagram* {
         (v1) -- [fermion] (i1),
         (v1) -- [fermion] (f1),
         (v1) -- [scalar,edge label' = $\tilde{b}$] (v2),
         (i2) -- [fermion] (v2),
        (f2) -- [fermion] (v2)
        };

       \end{feynman}
\end{tikzpicture}
    \end{subfigure}
    \hspace{0.15\linewidth}
    \begin{subfigure}{0.13\linewidth}
        \centering
         \begin{tikzpicture}
            \begin{feynman}
        \vertex (i1) at (-2.3, 1.3) {${d}$} ;
        \vertex (i2) at (-2.3, -1.3) {${g}$} ; 
        \vertex (f1) at (-1, 0) ;
        \vertex (f2) at (0, 0) ;
        \vertex (f3) at (1, 0.9) ;
        \vertex (v1) at (1.6, -1.1) {$\bar{t}$};
        \vertex (v2) at (2.3, 1.3) {${d}$};
        \vertex (v3) at (2.3, 0.5) {${t}$};
        \diagram* {
            (i1)  -- [fermion] (f1) -- [gluon] (i2),
            (f1) -- [fermion] (f2), 
            (v1)-- [fermion] (f2) -- [scalar,edge label' = $\tilde{b}^*$] (f3), 
          (f3) -- [fermion] (v2),
          (f3) -- [fermion] (v3)
        };

    \end{feynman}
\end{tikzpicture}
\end{subfigure}
       \hspace{0.23\linewidth}
  \begin{subfigure}{0.3\linewidth}
        \centering
         \begin{tikzpicture}
            \begin{feynman}
       \vertex (i1) at (-1.8, 1.3) {$d$};
        \vertex (i2) at (-1.8,-1.3) {$g$};
         \vertex (f1) at (0, 0.9) ;
        \vertex (f2) at (1.0,0.9);
         \vertex (f3) at (0,-0.9);
        \vertex (v1) at (2.3,1.3){$d$};
         \vertex (v2) at (2.3,0.5){$t$};
         \vertex (v3) at (2.3,-1.3){$\bar{t}$};
     \diagram* {
            (i1)  -- [fermion] (f1),
           (f1) -- [scalar, edge label' = $\tilde{b}^*$] (f2),
          (f2) -- [fermion] (v1),
          (f2) -- [fermion] (v2),
           (i2)  -- [gluon] (f3),
           (f3)  -- [fermion] (f1),
           (v3)  -- [fermion] (f3)
        };
    
    \end{feynman}
\end{tikzpicture}
\end{subfigure}     
 
\quad

  \begin{subfigure}{0.13\linewidth}
        \centering
         \begin{tikzpicture}
            \begin{feynman}
       \vertex (i1) at (-1.8, 1.3) {$d$};
        \vertex (i2) at (-1.8,-1.3) {$g$};
         \vertex (f1) at (0, -0.9) ;
        \vertex (f2) at (1.0,-0.9);
         \vertex (f3) at (0,0.9);
        \vertex (v1) at (2.1,-1.3){$d$};
         \vertex (v2) at (2.1,-0.5){$t$};
         \vertex (v3) at (2.1,1.3){$\bar{t}$};
     \diagram* {
            (i2)  -- [gluon] (f1),
           (f1) -- [scalar, edge label' = $\tilde{b}^*$] (f2),
          (f2) -- [fermion] (v1),
          (f2) -- [fermion] (v2),
           (i1)  -- [fermion] (f3),
           (f3)  -- [scalar, edge label' = $\tilde{b}$] (f1),
           (v3)  -- [fermion] (f3)
        };
    
    \end{feynman}
\end{tikzpicture}
\end{subfigure}     
 \hspace{0.18\linewidth}
    \begin{subfigure}{0.13\linewidth}
        \centering
         \begin{tikzpicture}
            \begin{feynman}
       \vertex (i1) at (-1.8, 1.3) {$\bar{d}$};
        \vertex (i2) at (-1.8,-1.3) {$d$};
         \vertex (f1) at (0, 0.9) ;
        \vertex (f2) at (1.0,0.9);
         \vertex (f3) at (0,-0.9);
          \vertex (f4) at (1,-0.9);
        \vertex (v1) at (2.2,1.3){$\bar{d}$};
         \vertex (v2) at (2.2,0.5){$\bar{t}$};
         \vertex (v3) at (2.2,-1.3){${t}$};
         \vertex (v4) at (2.2,-0.5){${d}$};
     \diagram* {
            (f1)  -- [fermion] (i1),
           (f1) -- [scalar, edge label' = $\tilde{b}$] (f2),
          (v1) -- [fermion] (f2),
          (v2) -- [fermion] (f2),
           (i2)  -- [fermion] (f3),
           (f3)  -- [scalar, edge label' = $\tilde{b}^*$] (f4),
           (f4)  -- [fermion] (v3),
           (f4)  -- [fermion] (v4),
           (f1)  -- [fermion] (f3),
        };
    
    \end{feynman}
\end{tikzpicture}
\end{subfigure}     
 \hspace{0.2\linewidth}    
    \begin{subfigure}{0.3\linewidth}
        \centering
         \begin{tikzpicture}
            \begin{feynman}
       \vertex (i1) at (-1, 1.3) {$\bar{d}$};
        \vertex (i2) at (-1,-1.3) {$d$};
         \vertex (f1) at (0, 0) ;
        \vertex (f2) at (1.0,0);
         \vertex (f3) at (2,0.9);
          \vertex (f4) at (2,-0.9);
        \vertex (v1) at (3.2,1.3){$\bar{d}$};
         \vertex (v2) at (3.2,0.5){$\bar{t}$};
         \vertex (v3) at (3.2,-1.3){${t}$};
         \vertex (v4) at (3.2,-0.5){${d}$};
     \diagram* {
            (i2)  -- [fermion] (f1) --[fermion] (i1),
           (f1) -- [gluon] (f2),
          (f2) -- [scalar, edge label' = $\tilde{b}$] (f3),
          (f2) --  [scalar, edge label' = $\tilde{b}^*$] (f4),
           (v1)  -- [fermion] (f3),
           (v2)   -- [fermion] (f3),
           (f4)  -- [fermion] (v3),
           (f4)  -- [fermion] (v4),
         
        };
    
    \end{feynman}
\end{tikzpicture}
\end{subfigure}      

\quad

    \begin{subfigure}{0.3\linewidth}
        \centering
         \begin{tikzpicture}
            \begin{feynman}
       \vertex (i1) at (-1, 1.3) {$g$};
        \vertex (i2) at (-1,-1.3) {$g$};
         \vertex (f1) at (0, 0) ;
        \vertex (f2) at (1.0,0);
         \vertex (f3) at (2,0.9);
          \vertex (f4) at (2,-0.9);
        \vertex (v1) at (3.2,1.3){$\bar{d}$};
         \vertex (v2) at (3.2,0.5){$\bar{t}$};
         \vertex (v3) at (3.2,-1.3){${t}$};
         \vertex (v4) at (3.2,-0.5){${d}$};
     \diagram* {
            (i2)  -- [gluon] (f1) --[gluon] (i1),
           (f1) -- [gluon] (f2),
          (f2) -- [scalar, edge label' = $\tilde{b}$] (f3),
          (f2) --  [scalar, edge label' = $\tilde{b}^*$] (f4),
           (v1)  -- [fermion] (f3),
           (v2)   -- [fermion] (f3),
           (f4)  -- [fermion] (v3),
           (f4)  -- [fermion] (v4),
         
        };
    
    \end{feynman}
\end{tikzpicture}
\end{subfigure}      
\caption{Diagrams illustrating $t\bar{t}$ production (top left),
  $t\bar{t}j$ production from processes that can include a single
  on--shell anti--sbottom in the intermediate state (the next three
  diagrams), as well as $t\bar{t}jj$ production from processes which
  can include an on--shell $\tilde b_R \tilde b_R^*$ pair (the last
  three diagrams). Contributions with possible on--shell single
  sbottom, $p ~p \rightarrow \tilde{b}~ t\rightarrow t \bar{t}j$, are
  not shown but included in the simulation; the corresponding diagrams
  are identical to the second to fourth diagrams, except that all
  particles are replaced by their corresponding
  anti--particles. Additional diagrams contributing to $t \bar t j$ or
  $t \bar t jj$ production where no intermediate (anti--)sbottom can
  become on--shell represent higher--order contributions and are
  therefore not included in our simulation.}
\label{fig:feyn-diagram}
\end{figure}
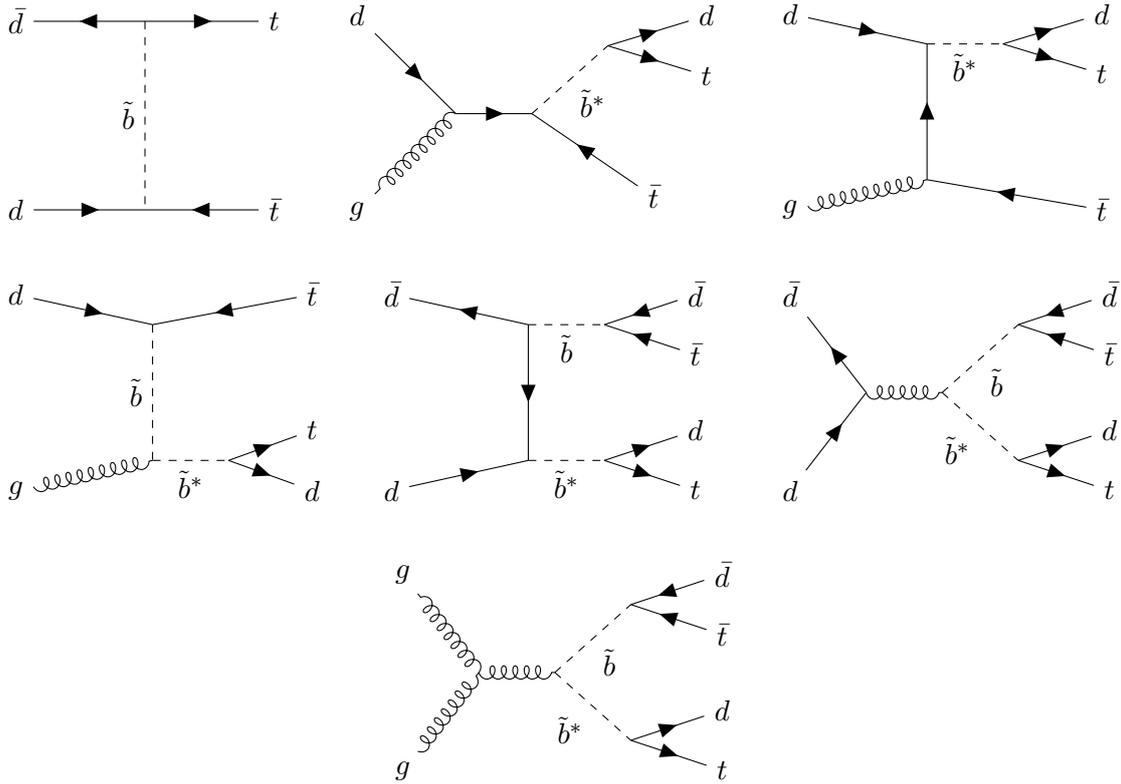

As shown in Fig.~\ref{fig:feyn-diagram}, we include three classes of
contributions to inclusive top pair production due to RPV
interactions. The first diagram shows direct $t\bar{t}$ production via
sbottom exchange in the $t-$channel. The next three diagrams show
$t\bar{t}j$ production with a single, possibly on--shell,
anti--sbottom in the intermediate state; we also include the charge
conjugate diagrams (not shown) in our simulation. The last three
diagrams show $t\bar{t}jj$ production with a possibly on--shell
sbottom pair in the intermediate state. The first diagram is of order
$\left|\lambda^{\prime\prime}_{313}\right|^2$. If all sbottoms are
off--shell in the $t\bar{t}j$ and $t\bar{t}jj$ channels, all diagrams
except the first one are higher order corrections. However, this is
not true if the sbottoms are on--shell. In this case, the second to
fourth diagrams are of order $g_S \lambda^{\prime\prime}_{313}$ and
the last three diagrams are of order $g^2_S $, where $g_S$ is the
$SU(3)_c$ coupling constant. We include both on-- and off--shell
sbottom exchange in these diagrams via the use of Breit--Wigner
propagators for the sbottom squarks carrying time--like
$4-$momentum. Although other diagrams exist, they only involve the
exchange of off--shell sbottoms, resulting in higher order
contributions. Therefore, they are not included in the simulations.

In the SMEFT framework, the sbottom squarks shown in
Fig.~\ref{fig:feyn-diagram} should be integrated out. The first
diagram is then described by the $4-$quark operators of
eqs.(\ref{wilson}); it contributes in leading order. In this language
the second and third diagrams show a subset of higher order
contributions where a gluon is attached to one of the quarks
participating in these $4-$quark operators. The fourth, sixth and
seventh diagrams contain two sbottom propagators coupling to a gluon
and therefore do not appear in the SMEFT at dimension 6. Finally, the
fifth diagram can be obtained by two $4-$quark operators together via
the exchange of a top quark. It thus needs two SMEFT vertices, which
means it can be considered a higher order correction to inclusive top
pair production in the SMEFT language; we just saw that this is true
also in the RPV model, if (and only if) the sbottom squarks are
off--shell.  In the SMEFT simulation we therefore only include the
leading contribution corresponding to the first diagram. The NLO
correction to the $t\bar{t}$ production is small as explained in
ref.~\cite{Degrande:2020evl}, which finds $K-$factors
($ K = \sigma_{NLO}/\sigma_{LO}$) close to unity for the operators
$\mathcal{O}_{t d}^{(1)}$ and $\mathcal{O}_{t d}^{(8)}$.

We perform our simulation using \textbf{MadGraph5$\_$aMC@NLO}
\cite{Alwall:2014hca,Frederix:2018nkq} to generate $500k$
parton--level inclusive $t\bar{t}$ events. We use the parton
distributions function (PDF) set NNPDF23$\_$nlo$\_$as$\_$0119, with
factorization and renormalization scales set to
$\mu_R=\mu_F=\frac{1}{2}(\sqrt{m_t^2+p_T^2(t)}+\sqrt{m_t^2+p_T^2(\bar{t})}
)$. The squared Feynman amplitude is expressed as
\begin{equation} \label{amp}
  |\mathcal{M}|^2 = |\mathcal{M}_{\text{SM}}|^2 + 2 \text{Re}(
  \mathcal{M}^*_{\text{SM}} \mathcal{M}_{\text{BSM}})
  + |\mathcal{M}_{\text{BSM}}|^2\,.
\end{equation}
As explained in the Introduction, the interference term will be called
``linear RPV (or EFT)'' and the last term will be called ``quadratic
RPV (or EFT)'' in the following. The linear (quadratic) EFT term is
also called $\Lambda^{-2}$ ($\Lambda^{-4}$) term, since it is
suppressed by two (four) inverse powers of the large energy scale
$\Lambda$.

In order to compute $95\%$ confidence level (CL) limits we use the
$\chi^2$ statistic, requiring $\Delta\chi^2=3.84$. Here
$\chi^2 = \sum_{i,j} V_i M^{-1}_{ij} V_j$ where the $V_i$ are the
differences between measured and predicted values of a given
observable in the $i-$th bin, and $M$ is the covariance matrix that
includes both statistical and systematic uncertainties from the
experiment, as well as the estimated theory uncertainties from NNLO SM
simulation. Both the NNLO SM predictions of the distributions and the
complete covariance matrices are provided by the CMS
\cite{CMS:2021vhb} and ATLAS \cite{ATLAS:2022waa}
collaboration. Finally, $\Delta \chi^2$ is the difference between the
value of $\chi^2$ computed for some set of BSM parameters (SMEFT
Wilson coefficients, or RPV coupling and sbottom mass) and the SM
value of $\chi^2$. This is conservative in that it gives relatively
weaker bounds than computing $\Delta \chi^2$ relative to the BSM set
of parameters that minimizes the value of $\chi^2$; in a BSM theory
this minimal $\chi^2$ can be below the SM prediction, but cannot be
above it if very small new couplings are allowed.

\subsection{Resonance peak in the $t\bar{t}j$ production }
\label{sec3.2}

\begin{figure}[htbp]
\centering
\includegraphics[scale=0.5]{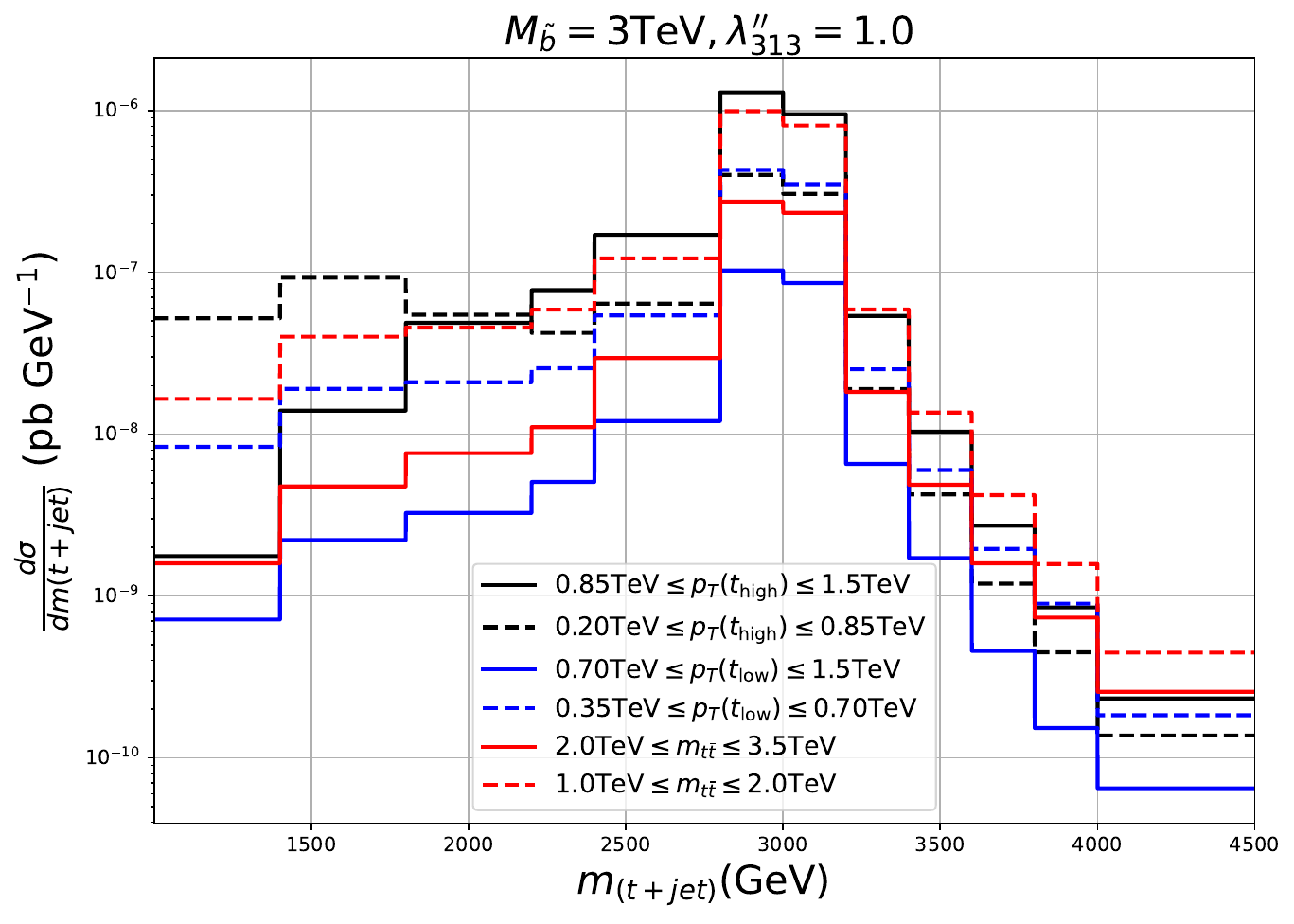}
\caption{The top + jet invariant mass distributions for the
  $p ~p \rightarrow \tilde{b}^*~ \bar{t}\rightarrow t~ \bar{t}~ j$
  process as described by the 2nd to 4th diagrams of
  Fig.~\ref{fig:feyn-diagram}, for a sbottom mass of $3$ TeV. We have
  used a Breit--Wigner propagator for the unstable anti-sbottom, with
  decay width given by eq.(\ref{decay}). The histograms show results
  after imposing the indicated kinematic cuts. Here
  $p_T(t_{\rm high})$ and $p_T(t_{\rm low})$ are the higher and lower
  transverse momentum of the top or anti-top quark, respectively, and
  $m_{t\bar{t}}$ is the invariant mass of the $t \bar t$ pair.}
\label{fig:resonance_dis}
\end{figure}

As mentioned above, in the $t\bar{t}j$ channel, only diagrams
corresponding to single sbottom or antisbottom production (shown by
the 2nd to 4th diagrams of Fig.~\ref{fig:feyn-diagram} and their
charge conjugate versions) are considered.
Fig.~\ref{fig:resonance_dis} shows the resulting top + jet invariant
mass distribution for $M_{\tilde{b}}=3$ TeV for the process
$p ~p \rightarrow \tilde{b}^*~ \bar{t}\rightarrow t \bar{t}j$; the
charge conjugate process requires a $\bar d$ quark in the initial
state and therefore has a considerably smaller cross section. We have
imposed additional cuts on a quantity characterizing the ``hardness''
of the process: the higher [$p_T(t_{\rm high})$] and lower
[$p_T(t_{\rm low})$] of the transverse momenta of the $t$ and
$\bar t$, or the $t \bar t$ invariant mass. We see that a prominent
peak at $3$ TeV, the mass of the produced $\tilde b_R^*$ squark,
remains even if we artificially remove very hard events.

Of course, this peak is not reproduced by the SMEFT. Hence the upper
bound on the coupling $\lambda^{\prime\prime}_{313}$ derived in the
SMEFT framework is likely to differ significantly from that derived in
the RPV model for sbottom mass up to at least $3$ TeV if the
contribution of the $t\bar{t}j$ channel remains sizable. For
sufficiently large $M_{\tilde b}$ the resonance peak will certainly
disappear; however, the total sbottom exchange contribution will also
become smaller with increasing sbottom mass, hence very large
$M_{\tilde b}$ will not lead to measurable effects in current LHC
data. Alternatively, one could impose an upper cut directly on the
invariant mass of the top + jet system, thereby removing the resonance
peak. While this would reduce the difference between the SMEFT and RPV
results, it would also remove many signal events, and would thus lead
to an artificially weakened constraint on the RPV model.

The last three diagrams shown in Fig.~\ref{fig:feyn-diagram}
contribute to $t\bar{t}jj$ production within the RPV model via the
production of a $\tilde b_R \tilde b_R^*$ pair. Contributions where at
least one of them is on--shell again cannot be described by the SMEFT.
However, we will see below that for sbottom masses of interest these
diagrams contribute little to inclusive $t \bar t$ production.

In the following subsections, we will present distributions of the
observables that have been measured by CMS or ATLAS; these
measurements can be used to constrain the original RPV model or its
implementation in the SMEFT using eqs.(\ref{wilson1}). Recall that our
main goal is to check whether the latter gives a faithful
representation of the former as far as current LHC data are
concerned. The discussion above indicates that this is only true if
the contribution from the $t\bar{t}j$ channel to a given distribution
is small even in the RPV model {\em and} if the RPV model and SMEFT
predict similar distributions for the $t\bar{t}$ channel. We will
compare the RPV and SMEFT predictions for the relevant distributions
assuming $\lambda^{\prime\prime}_{313} = 1$, so that the theory
remains perturbative up to very high scales. We will also compare the
resulting exclusion limits derived in the two frameworks; here we will
only require $\lambda^{\prime\prime}_{313}<4.0$ so that the sbottom
decay width remains below its mass.

\subsection{Transverse momentum of top quarks}

\begin{figure}[htbp]
\centering
\includegraphics[scale=0.38]{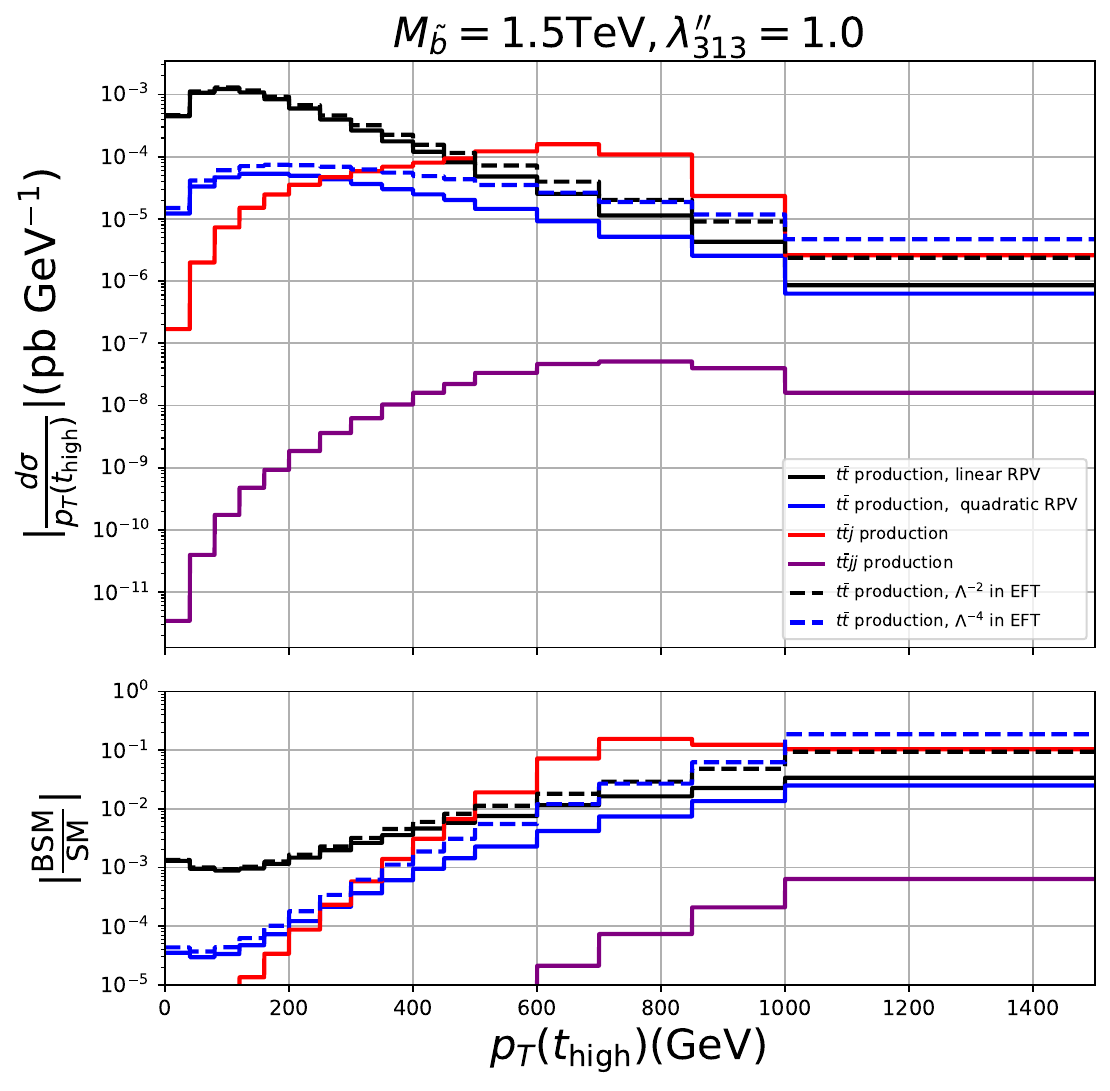}
\hspace{0.1in}
\includegraphics[scale=0.38]{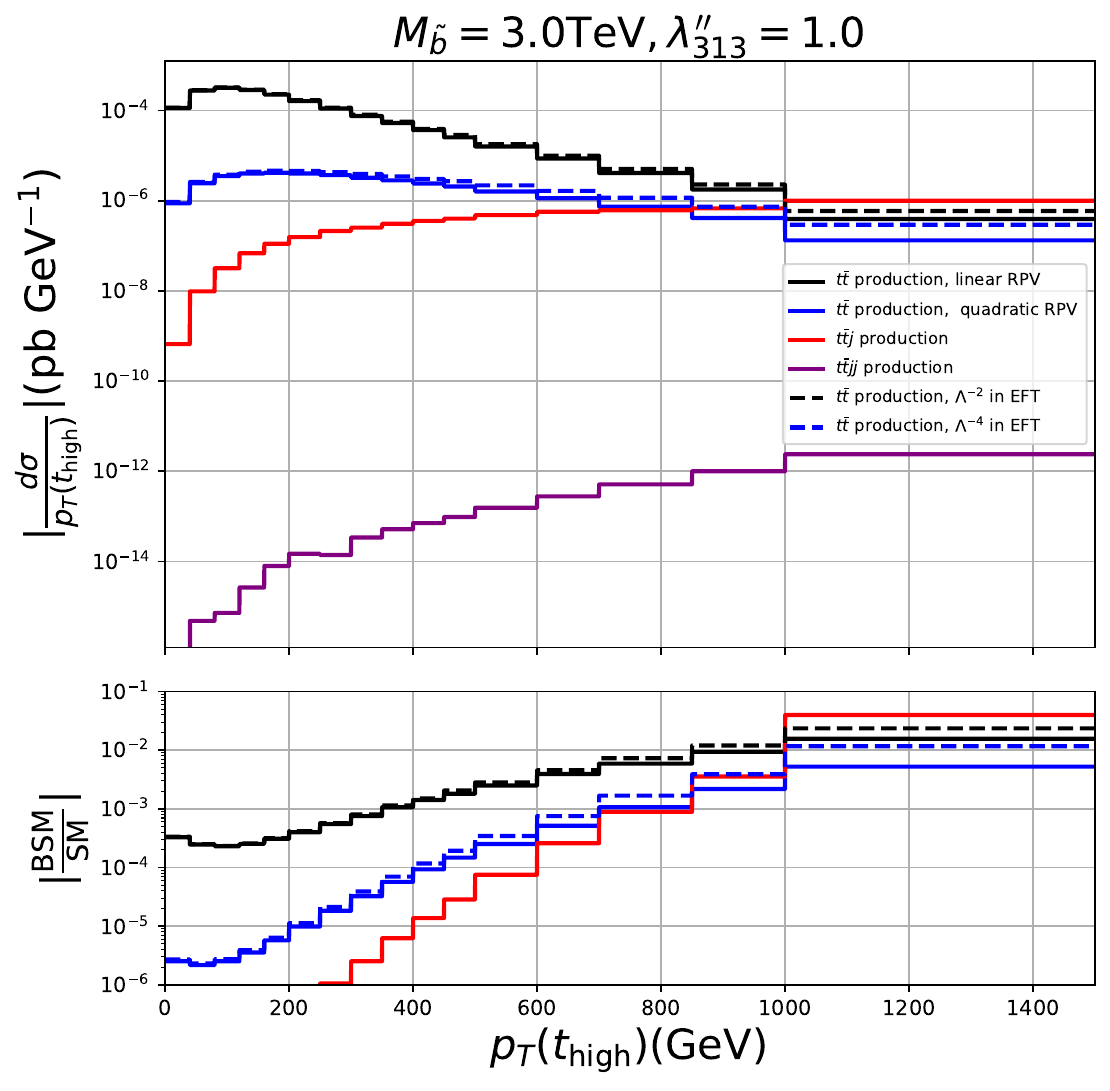}
\caption{Distributions of $p_T(t_{\rm high})$ at parton level
  predicted by the RPV model (solid) and its SMEFT implementation
  (dashed). The black histograms show the absolute value of the
  (negative) interference terms with the QCD diagrams, i.e. the linear
  new physics contributions to the exclusive $t\bar{t}$ channel. The
  blue histograms show the square of the $\tilde b_R$ exchange
  contribution to the $t \bar t$ channel, i.e. the quadratic new
  physics contributions. The red and violet histograms show the
  (purely new physics, i.e. quadratic) contributions to the
  $t \bar t j$ and $t \bar t jj$ channels, respectively; here only the
  RPV model contributes to leading order. The left (right) frames are
  for a sbottom mass of $1.5$ ($3.0$) TeV. The top frames show the
  absolute distributions, whereas the lower frames show results
  normalized to the NNLO QCD prediction.}
\label{fig:pthigh_dis}
\end{figure}

In this subsection, we analyze the transverse momentum distributions
of the produced top (anti--)quarks. We begin with
Fig.~\ref{fig:pthigh_dis}, which presents the distributions of the
larger of the two $p_T$ values, $p_T(t_{\rm high})$; only the BSM
contributions are shown. The black histograms show the absolute value
of the (negative) interference term with the QCD diagrams. The
difference between the SMEFT (dashed) and full RPV model (solid)
predictions becomes sizable at
$p_T(t_{\rm high}) \simeq M_{\tilde b}/3$. Even though this
contribution peaks in absolute value at about half the top mass, it
falls off less quickly than the pure SM prediction does, i.e. the new
contribution becomes more significant at larger $p_T(t_{\rm high})$,
as shown in the lower frames.

The blue histograms show the squared sbottom exchange contribution to
exclusive $t \bar t$ production. This contribution, which is ignored
in linear SMEFT (or RPV) fits, becomes at least comparable in
magnitude to the interference term for $p_T(t_{\rm high}) \geq 1$
TeV. Moreover, since this contribution contains the square of the
sbottom propagator, the difference between the RPV and SMEFT
predictions is larger than for the interference term. In fact, the
SMEFT predicts that for $M_{\tilde b} = 1.5$ TeV (left frames) this
term dominates over the interference term for
$p_T(t_{\rm high}) \geq 1$ TeV, which does not happen in the full RPV
calculation.

Note that the SMEFT prediction always exceeds that of the full RPV
model, both in linear and in quadratic order. This can be understood
from the observation that a space--like momentum flows through the
$\tilde b_R$ propagator in the first diagram shown in
Fig.~\ref{fig:feyn-diagram}; neglecting the momentum dependence of
this propagator, as done in the SMEFT, therefore over--estimates its
absolute value.

The red and magenta histograms show contributions from the
$t \bar t j$ and $t \bar t jj$ channels; recall that only the RPV
model contributes here at leading order, via the production of one or
two on--shell $\tilde b_R$ (anti--)squarks. We see that for
$M_{\tilde b} = 1.5$ TeV single sbottom production dominates the total
RPV contribution for $p_T(t_{\rm high}) \geq M_{\tilde b}/3$. Since
the SMEFT treatment predicts too large a contribution at large
$p_T(t_{\rm high})$, for $M_{\tilde b} = 1.5$ TeV the production of
on--shell sbottom squarks coincidentally improves the agreement
between the two predictions in this region. However, for
$M_{\tilde b} = 1.5$ TeV on--shell sbottom production also leads to a
pronounced (Jacobian) peak at
$p_T(t_{\rm high}) \simeq M_{\tilde b}/2$, where the harder top
(anti--)quark predominantly results from the decay of an on--shell
$\tilde b_R$; this is of course not reproduced by the SMEFT
prediction. Finally, we see that the $t \bar t jj$ contribution, due
to $\tilde b_R \tilde b_R^*$ pair production, remains well below the
other contributions even for $M_{\tilde b} = 1.5$ TeV, and is
completely negligible for $M_{\tilde b} = 3$ TeV.
 
\begin{table}[htbp]
\centering
\caption{Values of $\frac{\sigma(\text{SMEFT})}{\sigma(\text{RPV})}$ in
  two bins with moderate and high $p_T(t_{\rm high})$ and for two
  values of the sbottom mass, assuming
  $\lambda^{\prime\prime}_{313} = 1$. The third, fourth and fifth
  column only refer to the exclusive $t \bar t$ channel, showing the
  ratios of the linear terms, of the quadratic terms, and of the sums
  of linear and quadratic terms; in this channel
  $p_T(t_{\rm high}) = p_T(t_{\rm low})$ after correcting to the
  parton level. The last column shows the ratios for the fully
  inclusive $t \bar t$ cross section, including linear and quadratic
  contributions and the $t\bar t j$ and $t \bar t jj$ channels.}
\label{rpveft}
\begin{tabular}{|c|c|c|c|c|c|}
\hline
  $M_{\tilde{b}}$(GeV) & $p_T(t_{\text{high}})$ [GeV] &  $t \bar t$ (linear)
  &  $t \bar t$ (quadratic)  & $t \bar t$ (linear+quadratic) & total  \\
\hline
\multirow{2}{*}{1500}&500-600&1.50&2.42&1.10 & -0.42\\
\cline{2-6}
&1000-1500&2.77&7.46&-10.3 & 0.97 \\
\hline
\multirow{2}{*}{3000}&500-600&1.13&1.37&1.10 & 1.14 \\
\cline{2-6}
&1000-1500&1.50&2.23&1.14 & -0.41\\
\hline
\end{tabular}
\end{table}

The difference between the predictions by the RPV model and its SMEFT
implementation are summarized in in Table~\ref{rpveft}, for different
bins of $p_T(t_{\rm high})$. As noted above, the difference between
the SMEFT and full RPV predictions are larger for smaller sbottom
mass, for larger $p_T(t_{\rm high})$, and in quadratic (rather than
linear) order. Since the interference term is negative while the
squared BSM contribution is (of course) positive, the difference
between the SMEFT and RPV predictions can be significantly smaller in
the sum of linear and quadratic contributions to exclusive $t \bar t$
production than it is for either one separately. However, this very
same cancellation can also make the full RPV contribution to exclusive
$t \bar t$ production very small, as in the higher $p_T(t_{\rm high})$
bin for the lower sbottom mass; here the SMEFT predictions differs
from that of the (more) UV complete theory by an order of magnitude
and even gives the wrong sign! As already noted, in this particular
case including $t \bar t j$ production from (anti)sbottom decay in the
RPV happens to give close agreement between the two predictions for
inclusive $t \bar t$ production; however, doing so leads to very poor
agreement in the lower $p_T(t_{\rm high})$ bin.

For $M_{\tilde b} = 3$ TeV and $\lambda^{\prime\prime}_{313} = 1$ both
the SMEFT and the RPV model predict negative contributions to
exclusive $t \bar t$ production even in quadratic order; the 
predictions for the sum of the two contributions happen to agree over
a wide range of $p_T(t_{\rm high})$. Recall, however, that adding the
quadratic contribution in the SMEFT is not well motivated, since it is
of the same order in the cut--off parameter $\Lambda$ as contributions
from dimension$-8$ operators which are not included. Moreover, the last
column shows that this agreement is ruined once the $t \bar t j$
channel is included, which turns the total BSM contribution positive
at large $p_T(t_{\rm high})$ even for this large sbottom mass.

\begin{figure}[h]
\centering
\includegraphics[scale=0.5]{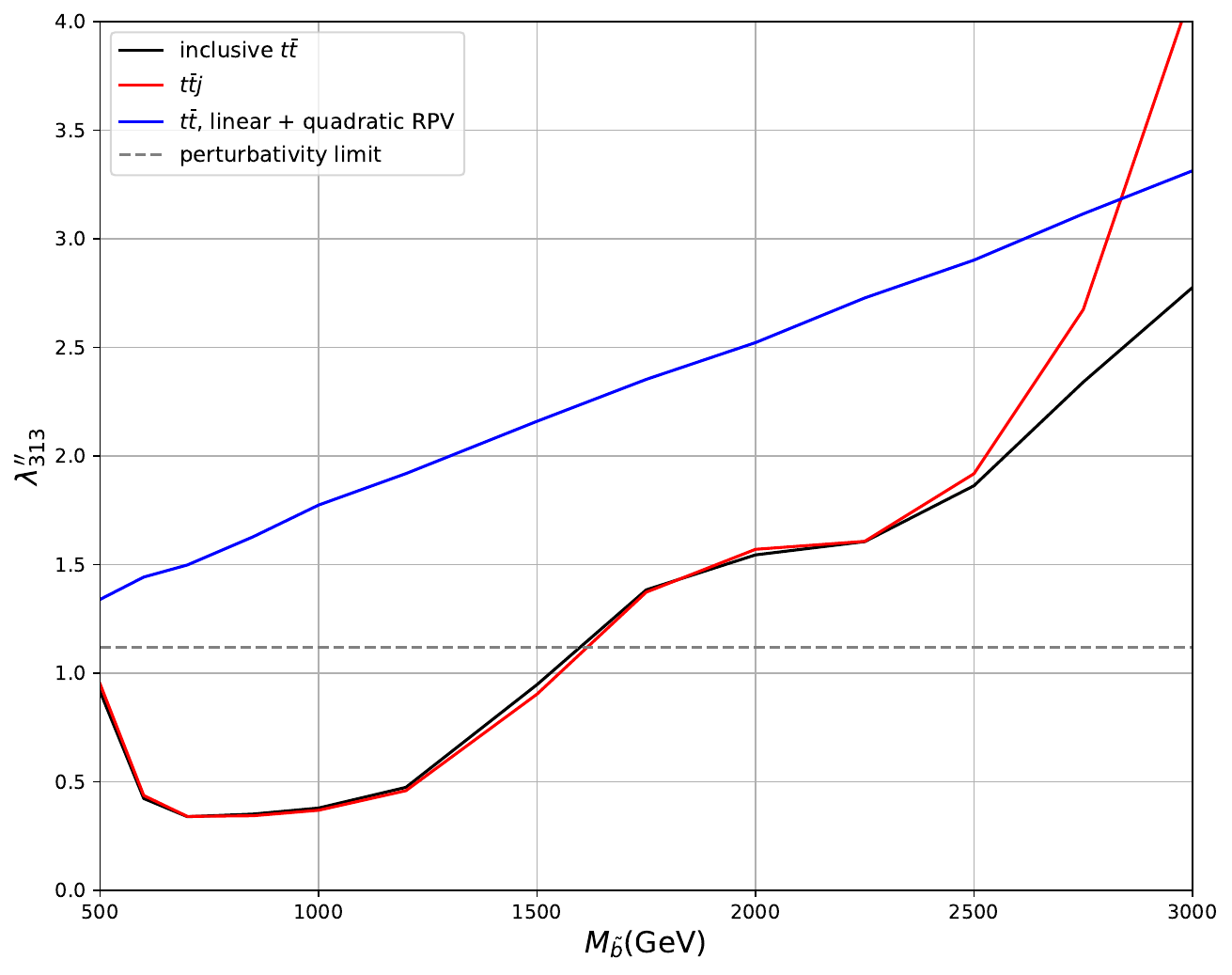}
\caption{Exclusion limits at $95\%$ confidence level, derived from the
  parton level distribution of $p_T(t_{\rm high})$ measured by CMS
  \cite{CMS:2021vhb}. The solid black line depicts the complete RPV
  bounds from inclusive $t \bar t$ production, where all Feynman
  diagrams shown in Fig.~\ref{fig:feyn-diagram} have been taken into
  account. The blue and red lines depict the bounds derived using only
  the sbottom contributions to the exclusive $t\bar{t}$ and
  $t\bar{t}j$ channel, respectively. The upper bound
  $\lambda^{\prime\prime}_{313} < 1.12$ is indicated by the dashed
  black line.}
\label{fig:pthigh_low_bound}
\end{figure}

Fig.~\ref{fig:pthigh_low_bound} presents the $95\%$ c.l. exclusion
limits derived by us from the CMS measurement \cite{CMS:2021vhb} of
the parton level $p_{T}(t_{\rm high})$ distribution. We see that for
$M_{\tilde b} \leq 2.5$ TeV the red curve, which has been derived
including $t \bar t j$ production only, nearly coincides with the
black line, where all diagrams depicted in Fig.~\ref{fig:feyn-diagram}
have been taken into account. Evidently the complete RPV bound is
primarily determined by single sbottom production leading to
$t\bar{t}j$ final states. This contribution remains significant even
for $M_{\tilde b} = 3$ TeV. Since it is not included in the SMEFT
treatment, bounds derived within the SMEFT framework will not be
reliable for any sbottom mass shown.

Considering even larger values of $M_{\tilde b}$ would, of course,
result in even weaker bounds on the coupling
$\lambda^{\prime\prime}_{313}$. Recall that this coupling should be
less than $1.12$ for the RPV model to be considered UV--complete,
since otherwise a Landau pole will appear well below the scale of
Grand Unification; if this bound is imposed, current data are not
sensitive to sbottom masses beyond $1.6$ TeV. Moreover, we saw in
Table~\ref{rpveft} that in the highest $p_T(t_{\rm high})$ bin the
quadratic SMEFT prediction exceeds that of the RPV model by more than
a factor of $2$ even for $M_{\tilde b} = 3$ TeV; this comparison is
independent of the value of $\lambda^{\prime\prime}_{313}$. The two
predictions will agree to within, say, $20\%$ only for $M_{\tilde b} > 5$
TeV. For these very large sbottom masses current data could only
exclude values of the coupling $\lambda^{\prime\prime}_{313}$ that
badly violate our perturbativity limit of $4$.

We finally note that in the full RPV model the bound on the coupling
does not increase monotonically with the sbottom mass. Of course, in
the SMEFT implementation only the ratio
$\lambda^{\prime\prime}_{313} / M_{\tilde b}$ can be constrained,
hence the bound on the coupling is necessarily proportional to the
mass. In our RPV model the bound on the coupling is strongest for
$M_{\tilde b} \simeq 750$ GeV. It becomes weaker for smaller sbottom
mass since the data actually slightly favor an RPV contribution with
$M_{\tilde b} \simeq 500$ GeV; however, the difference in $\chi^2$
relative to the SM value is not statistically significant.

\begin{figure}[htbp]
\centering
\includegraphics[scale=0.38]{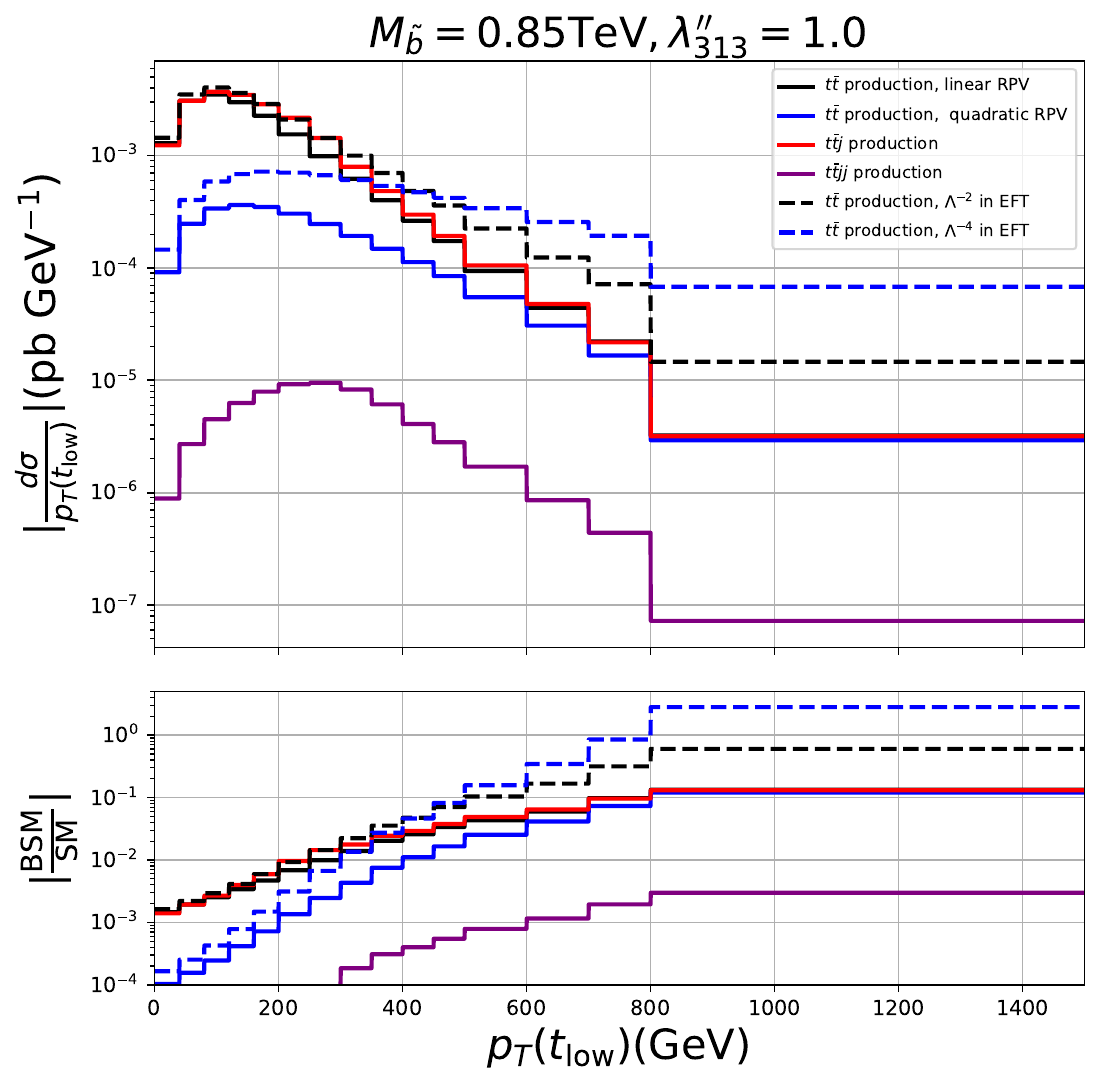}
\hspace{0.1in}
\includegraphics[scale=0.38]{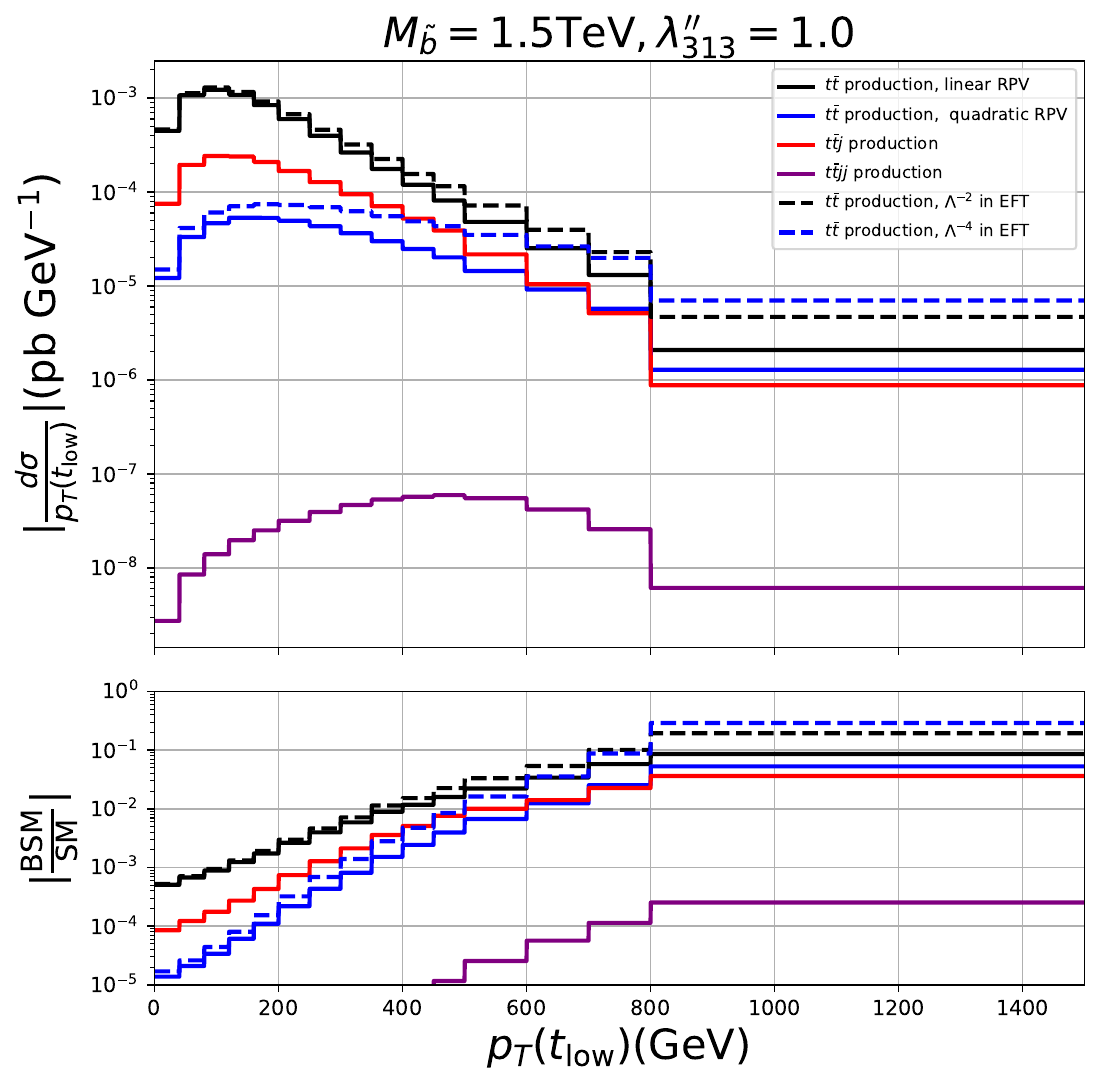}
\caption{As in Fig.\ref{fig:pthigh_dis}, but for the $p_T(t_{\rm low})$
  distribution and reduced values of $M_{\tilde b}$.}
\label{fig:ptlow_dis}
\end{figure}

We now turn to $p_T(t_{\rm low})$, the transverse momentum of the
softer of $t$ and $\bar t$; the corresponding distribution is shown in
Fig.~\ref{fig:ptlow_dis}. Here the contribution from the $t\bar{t}j$
channel, depicted by the red histograms, is not enhanced at
$p_T(t_{\rm low}) \sim \frac{M_{\tilde b}}{2}$, since in this channel
most of the time the softer top (anti-)quark is not produced from
squark decay. As a result this contribution is subdominant already for
$M_{\tilde b}=1.5$ TeV (right frame), in contrast to the
$p_T(t_{\rm high})$ distribution shown in the left frame of
Fig.~\ref{fig:pthigh_dis}. This reduces the difference between the RPV
and SMEFT predictions for the $p_T(t_{\rm low})$ distribution.

For $M_{\tilde{b}}=0.85$ TeV (left frame of Fig.~\ref{fig:ptlow_dis})
the $t \bar t j$ contribution is sizable in all bins, indicating that
the SMEFT description will not work. Recall that the interference
contribution depicted by the black histogram is negative; clearly in
the given example there's a strong cancellation between the new
contributions to the exclusive $t \bar t$ channel and the new
$t \bar t j$ channel, at least as far as the $p_T(t_{\rm low})$
distribution is concerned. Not surprisingly, for this smaller value of
$M_{\tilde{b}}$ the discrepancy between the SMEFT and full RPV
predictions for the exclusive $t \bar t$ channel becomes larger. For
example, the SMEFT predicts that the (positive) quadratic term exceeds
the (negative) linear one for $p_T(t_{\rm low}) > 500$ GeV; in the
full RPV this does not happen at all, at least not in the range of
transverse momenta shown. Note that in the exclusive $t\bar{t}$
channel at leading order the $p_{T}(t_{\rm high})$ and
$p_T(t_{\rm low})$ distributions are identical due to transverse
momentum conservation, i.e.,
$p_T(t_{\rm high}) \equiv p_T(t_{\rm low}) \equiv p_T(t) \equiv
p_T(\bar t)$.
  
\begin{figure}[h]
\centering
 \includegraphics[scale=0.5]{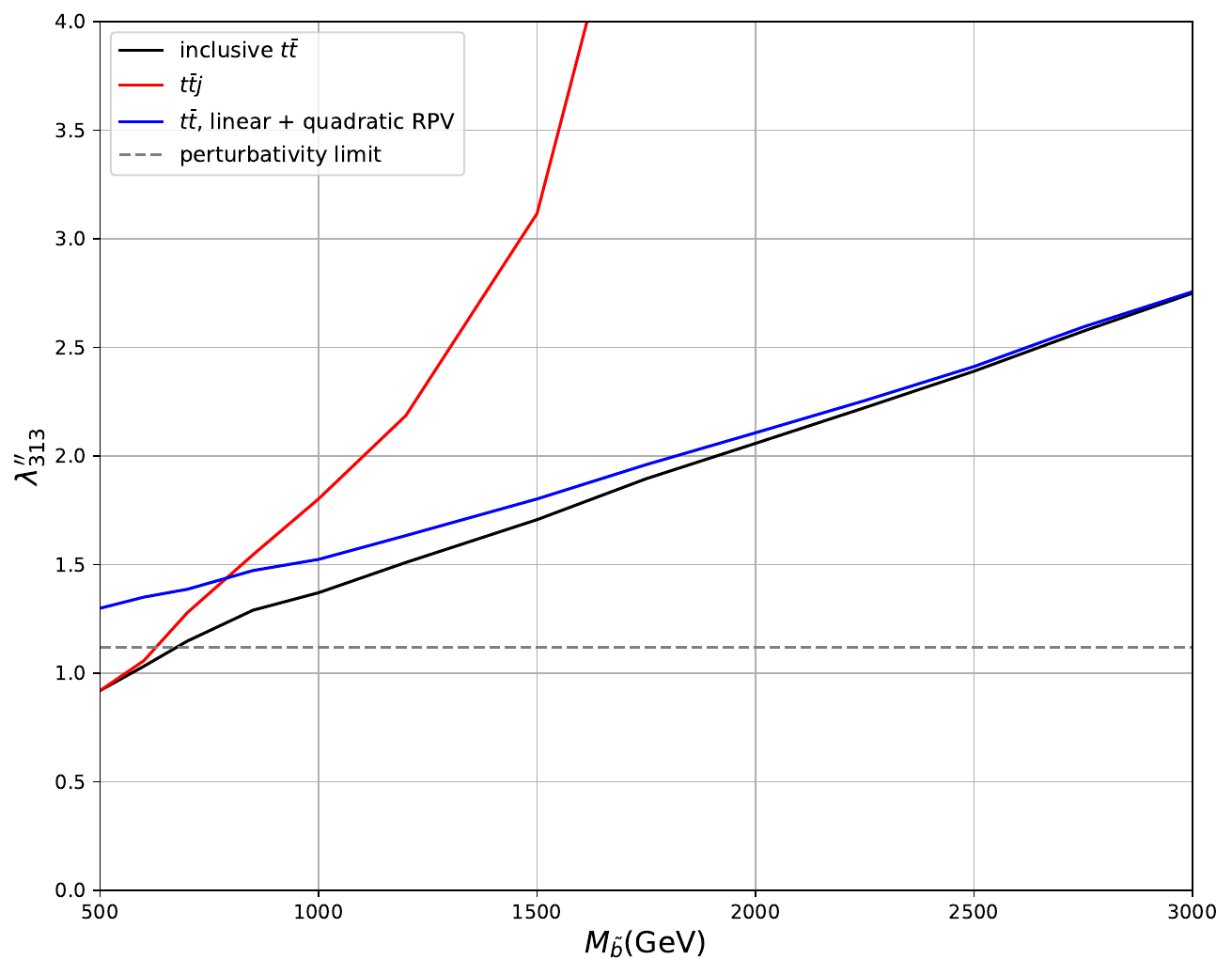}
 \caption{As in Fig.~\ref{fig:pthigh_low_bound}, but for exclusion
   limits derived from the $p_T(t_{\rm low})$ distribution.}
\label{fig:ptlow_bound}
\end{figure}

Since single $\tilde b_R$ production leading to the $t \bar t j$
channel is much less important in the $p_T(t_{\rm low})$ distribution
than in the $p_T(t_{\rm high})$ distribution, we expect the former to
lead to weaker bounds on the model parameters. This is confirmed by
Fig.~\ref{fig:ptlow_bound}, which presents the exclusion limits at
$95\%$ CL obtained from the CMS measurement \cite{CMS:2021vhb} of this
(parton--level) distribution. The black line, which shows the complete
RPV bound including all channels, now only coincides with the red line
derived from the $t \bar t j$ channel alone only for
$M_{\tilde b} \simeq 500$ GeV. For $M_{\tilde b} \geq 2$ TeV the black
line nearly coincides with the blue one, which has been obtained from
the exclusive $t \bar t$ channel alone, showing that the contribution
from the $t\bar{t}j$ channel becomes negligible here. Recall, however,
that the SMEFT does not reproduce the tail of the $p_T$ distribution
of the top (anti-)quark very well even for $M_{\tilde b} = 3$ TeV, see
Table~\ref{rpveft}. Moreover, current data for $p_T(t_{\rm low})$ lose
sensitivity to sbottom exchange or production once
$M_{\tilde b} > 680$ GeV if we demand that the theory remains
perturbative to very high energy scales.

\begin{figure}[h]
\centering
\includegraphics[scale=0.5]{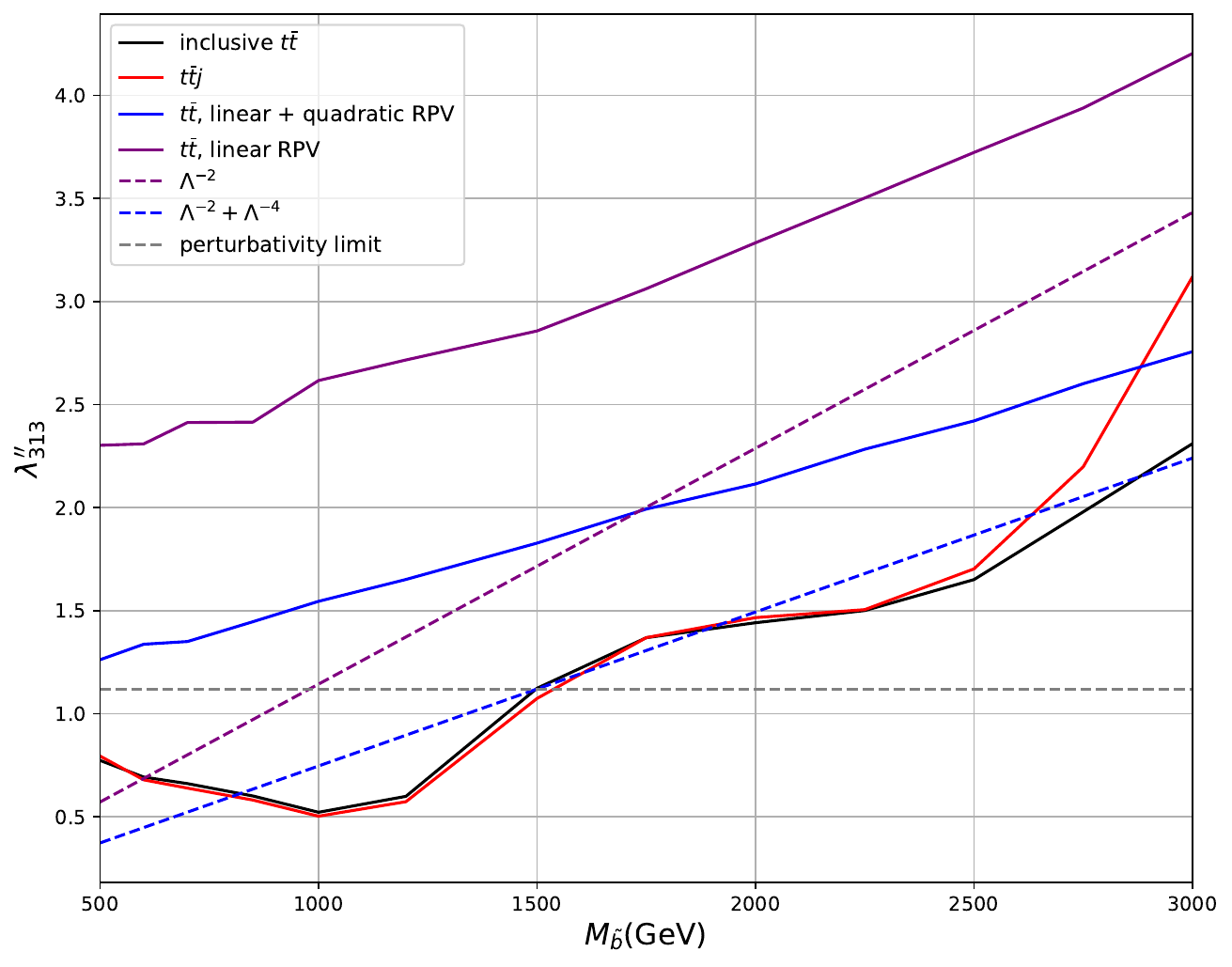}
\caption{Exclusion limits derived from the $p_T(t_h)$ distribution,
  where $t_h$ denotes the hadronically decaying top (anti-)quark. The
  solid purple line depicts the bound derived from the linear RPV
  contribution to the exclusive $t \bar t$ channel only; the meaning
  of the other solid lines is as in Fig.~\ref{fig:pthigh_low_bound}.
  The dashed purple and blue lines show the bounds derived from the
  SMEFT in linear and linear + quadratic order, respectively; they
  correspond to $\lambda^{\prime\prime}_{313}/M_{\tilde{b}}=1.14$ and
  $0.75$, respectively.}
\label{fig:ptbound}
\end{figure}

CMS also presents a measurement \cite{CMS:2021vhb} of the
parton--level $p_T(t_h)$ distribution, where $t_h$ denotes the
hadronically decaying top (anti-)quark, the other one decaying
semi--leptonically. Following Ref.~\cite{Catani:2019hip} we compute
the $p_T(t_h)$ distribution by averaging the $p_T$ distributions of
the top quark and antiquark; of course, in the exclusive $t\bar t$
channel we have $p_T(t_h) = p_T(t) = p_T(\bar t)$.

The resulting exclusion limits are shown in Fig.~\ref{fig:ptbound}.
The complete RPV bound (black solid line) is comparable to that
obtained from the $p_T(t_{\rm high})$ distribution (slightly weaker at
small $M_{\tilde b}$ and slightly stronger at larger $M_{\tilde b}$),
but stronger than that derived from the $p_{T}(t_{\rm low})$
distribution. It intersects the upper bound (\ref{uni-bound}) on
$\lambda^{\prime\prime}_{313}$ at $M_{\tilde{b}}\sim1.5$ TeV.

We also show the limit obtained by considering only the the exclusive
$t \bar t$ channel. The solid (dashed) purple and blue lines refer to
the bounds derived from linear RPV (SMEFT) and linear+quadratic RPV
(SMEFT) contributions, respectively. We first note that including the
square of the new contribution leads to much stronger constraints.
Even on the solid blue curve
$\lambda^{\prime\prime}_{313} / M_{\tilde b}$ is considerably larger
than in Figs.~\ref{fig:pthigh_dis} and \ref{fig:ptlow_dis}. This is
significant since the ratio of the square of the new contribution to
the interference term scales like
$\left(\lambda^{\prime\prime}_{313} / M_{\tilde b} \right)^2$, hence
along this line the quadratic terms are relatively much more important
than in our earlier example with $\lambda^{\prime\prime}_{313} = 1$.
On or near the bound the quadratic terms thus always dominate at large
$p_T(t_h)$.

By comparing lines of the same color, it is clear that the SMEFT
considerably overestimates the bounds, compared to those derived
within the UV complete RPV model. This agrees with our earlier
observation in Figs.~\ref{fig:pthigh_dis} and \ref{fig:ptlow_dis} and
Table~\ref{rpveft} that the SMEFT significantly overestimates the
squark exchange contribution even at $M_{\tilde b} = 3$ TeV,
especially in the bins with high $p_T$ where this contribution is most
significant. Evidently the quadratic SMEFT constraint very roughly reproduces
the complete RPV result. However, this is largely accidental, and also
misses that in the RPV model the bound on the coupling is not
strictly proportional to $M_{\tilde b}$. In particular, the bound gets
weaker at the smallest sbottom masses shown because the data again mildly
prefer a non--vanishing RPV contribution.\footnote{This should not be
  seen as independent evidence in favor of the RPV model, since the
  $p_T(t_{\rm high})$ and $p_T(t_h)$ distributions are derived from the
  same data set, and are hence strongly correlated.} We also remind the
reader that in the SMEFT framework including the square of $d=6$
operators but ignoring all $d=8$ operators is not really consistent.

\subsection{Invariant mass of the top pair}

\begin{figure}[htbp]
\centering
\includegraphics[scale=0.38]{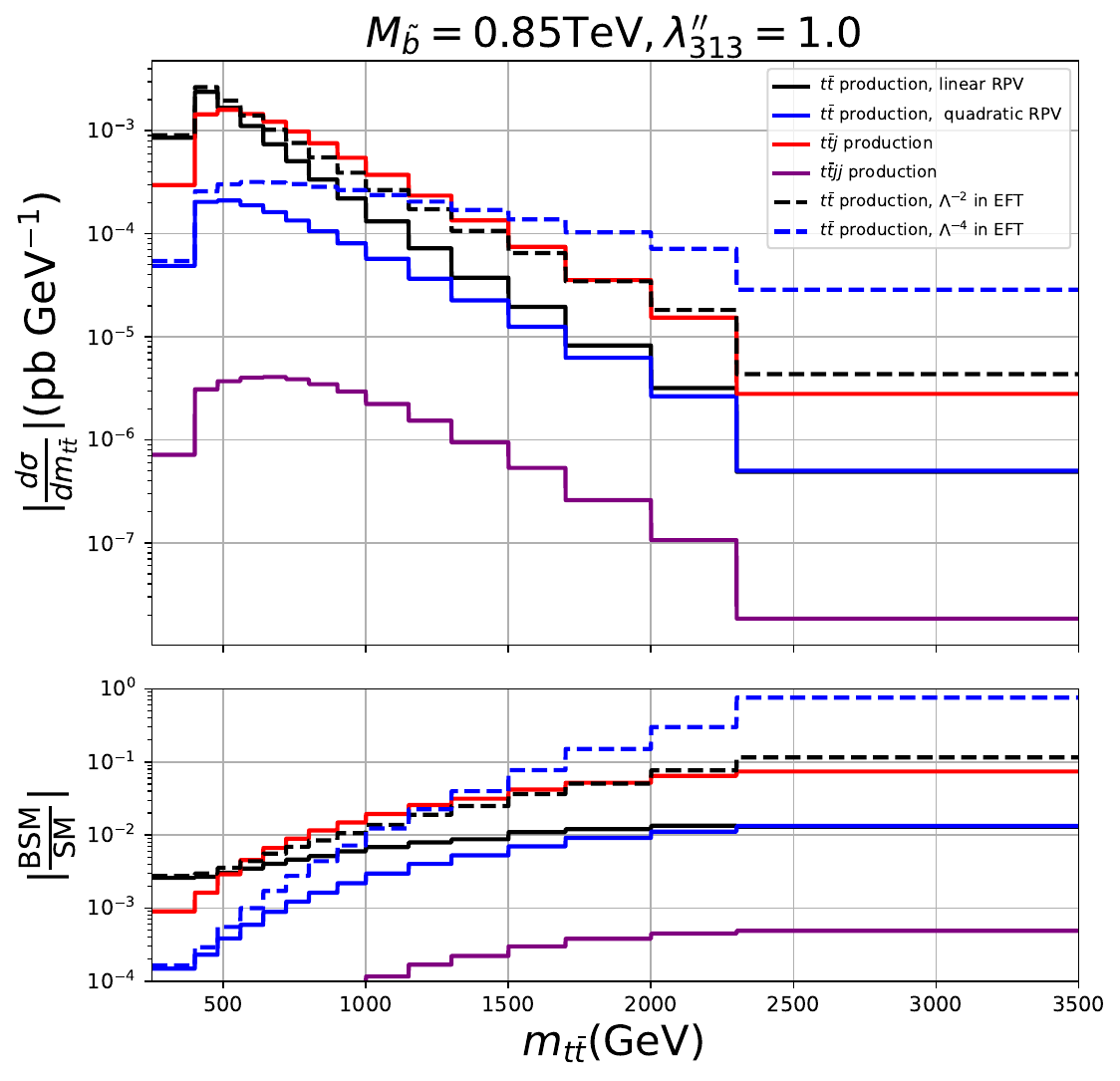}
\hspace{0.1in}
\includegraphics[scale=0.38]{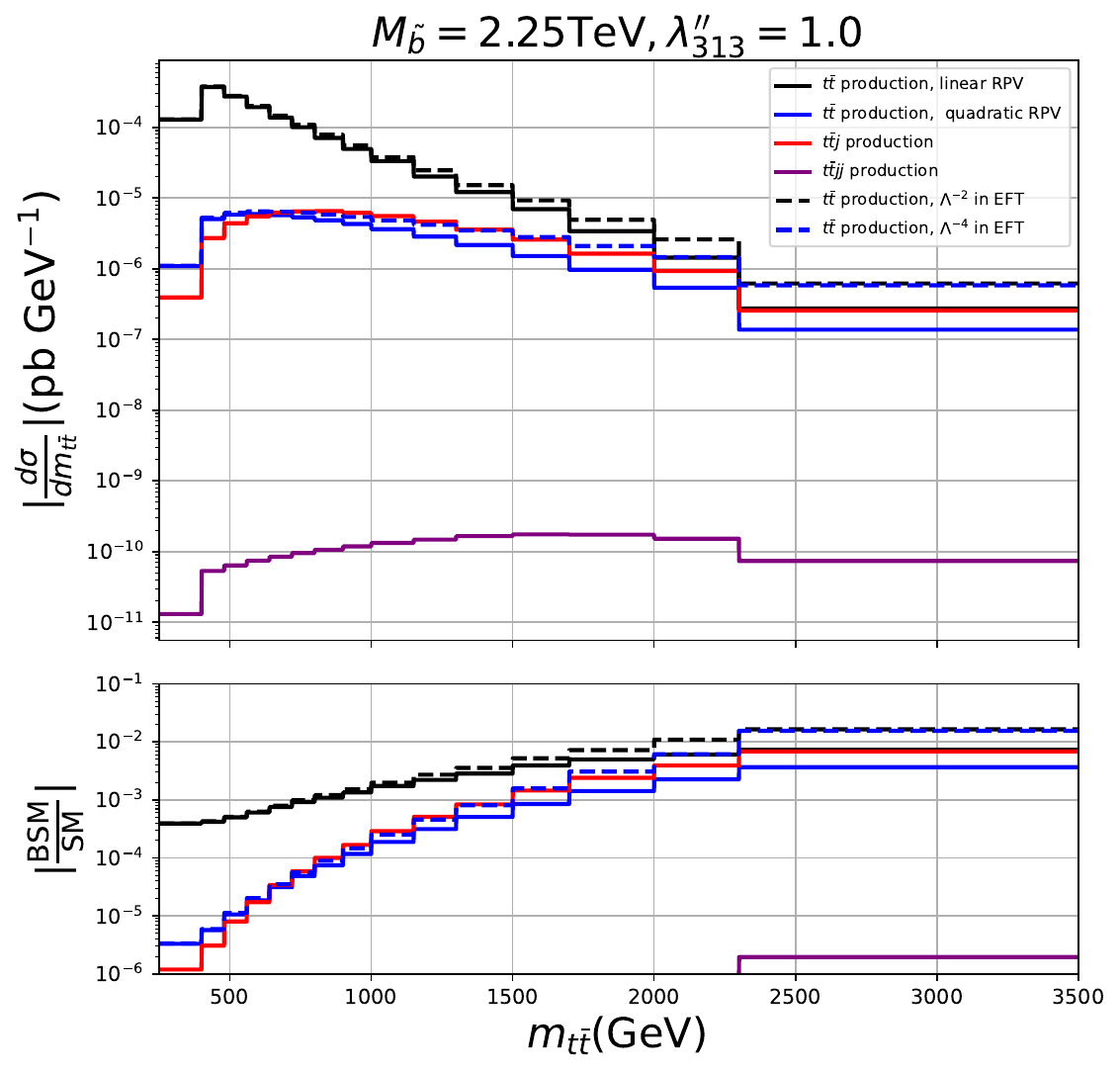}
\caption{As in Fig.~\ref{fig:pthigh_dis}, but for the $m_{t\bar{t}}$
  distribution.}
\label{fig:mtt_dis}
\end{figure}

We next turn to the distribution of the invariant mass of the
$t \bar t$ pair. Fig.~\ref{fig:mtt_dis} shows various BSM contributions
to this distributions derived from the RPV model (solid) and its SMEFT
implementation (dashed). As before, the lower frames show that the BSM
signal becomes more significant in the higher bins where the SM
contribution is strongly suppressed. The contribution from the
$t\bar{t}jj$ channel is again negligible in all bins.  Moreover, for
$M_{\tilde b} = 0.85$ TeV the $t\bar{t}j$ channel dominates for all
$m_{t \bar t} > 700$ GeV, while for $M_{\tilde b} = 2.25$ TeV this is
true in the last two bins. Moreover, the SMEFT again predicts too large
contributions to the exclusive $t \bar t$ channel. As before, the
discrepancy becomes worse for smaller $M_{\tilde b}$ and in the higher
bins, becoming significant for $m_{t\bar t} \gtrsim M_{\tilde b}$; it
is also worse in the quadratic terms than in the linear one.

\begin{table}[htbp]
\centering
\caption{Values of $\frac{\sigma(\text{SMEFT})}{\sigma(\text{RPV})}$ in
  two bins with moderate and high $m_{t \bar t}$ and for two
  values of the sbottom mass, assuming
  $\lambda^{\prime\prime}_{313} = 1$. The third, fourth and fifth
  column only refer to the exclusive $t \bar t$ channel, showing the ratios of
  the linear terms, of the quadratic terms, and of the sums of linear
  and quadratic terms. The last column shows the ratios for the fully
  inclusive $t \bar t$ cross section, including linear and quadratic
  contributions and the $t\bar t j$ and $t \bar t jj$ channels.}
\label{rpveft2}
\begin{tabular}{|c|c|c|c|c|c|}
\hline
  $M_{\tilde{b}}$(GeV) & $m_{t\bar t}$ [GeV] &  $t \bar t$ (linear)
  &  $t \bar t$ (quadratic)  & $t \bar t$ (linear+quadratic) & total  \\
\hline
\multirow{2}{*}{850}&720-800&1.50&2.26&1.23 & -0.74 \\
\cline{2-6}
&2300-3500&8.88&57.2&2050.04 & 8.62\\
\hline
\multirow{2}{*}{2250}&720-800&1.08&1.16&1.08 & 1.15\\
\cline{2-6}
&2300-3500&2.26&4.24&0.27 & -0.3 \\
\hline
\end{tabular}
\end{table}

Some results for the ratio $\sigma(\text{SMEFT})/\sigma(\text{RPV})$
are collected in Table~\ref{rpveft2}. The overall trends are quite
similar to those in Table~\ref{rpveft}. In particular, satisfactory
agreement is found only for $m^2_{t \bar t} \ll M^2_{\tilde b}$, as in
the penultimate row of Table~\ref{rpveft2}; otherwise even the sign of
the BSM contribution to inclusive $t \bar t$ production might be
predicted incorrectly by the SMEFT.

\begin{figure}[h]
\centering
\includegraphics[scale=0.5]{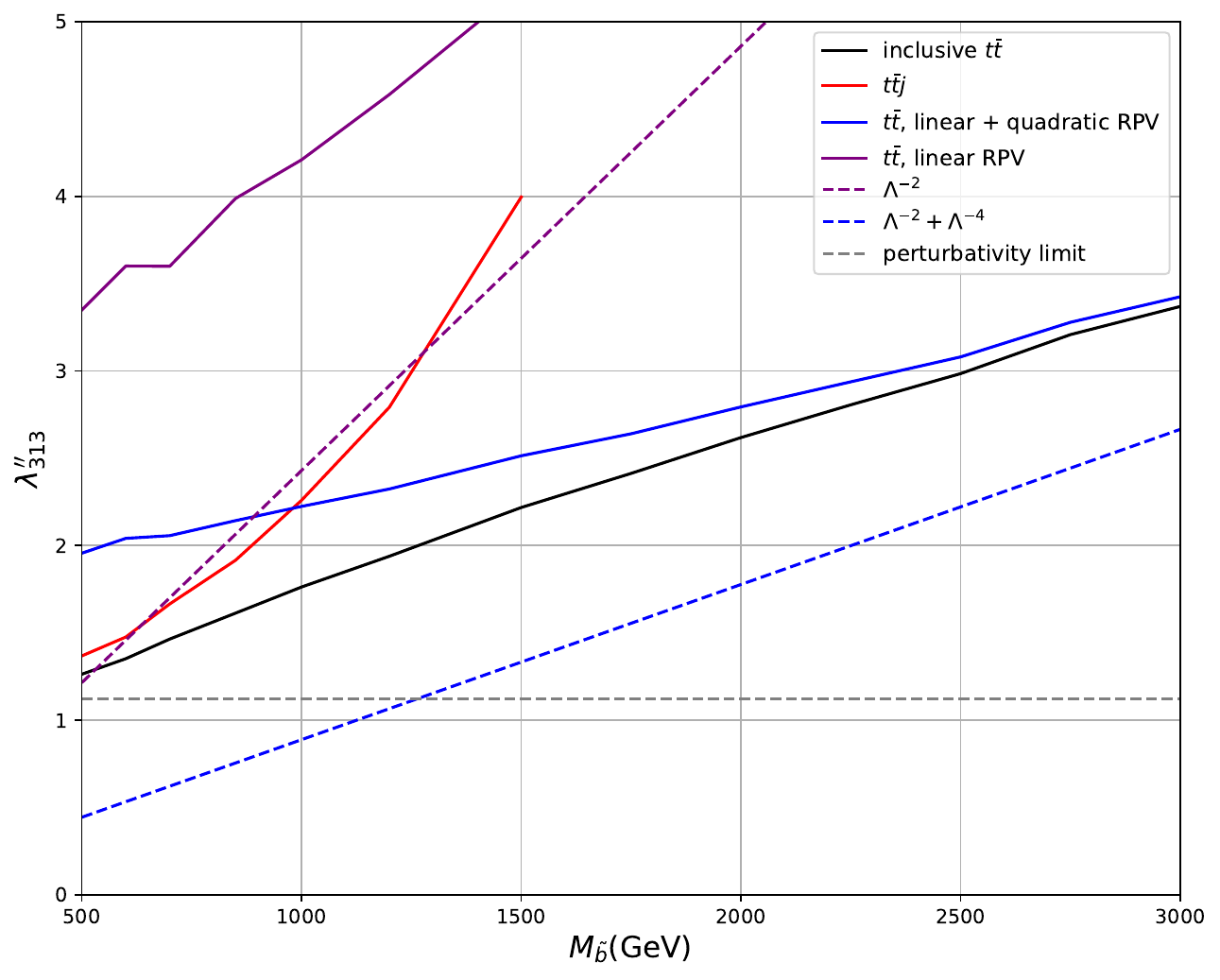}
\caption{As in Fig.~\ref{fig:pthigh_low_bound}, but for exclusion
  limits derived from the $m_{t\bar t}$ distribution measured by CMS.
  The dashed purple and blue lines correspond to
  $\lambda^{\prime\prime}_{313}/M_{\tilde{b}}=2.43$ and $0.89$ from
  including only linear and linear + quadratic SMEFT terms in the
  computation, respectively.}
\label{fig:mtt_bound}
\end{figure}

Fig.~\ref{fig:mtt_bound} presents the exclusion limits at $95\%$ CL
which we derive from the CMS measurement \cite{CMS:2021vhb} of the
parton--level $m_{t \bar t}$ distribution, again using a $\chi^2$ fit.
The complete RPV bound depicted by the black line is weaker than those
derived from the $p_T(t_{\rm high})$ or $p_T(t_h)$ distributions,
where on--shell single $\tilde b_R$ production led to a pronounced
(Jacobian) peak in the distribution. In the case at hand the BSM
contribution is more widely distributed. Moreover, large values of
$m_{t \bar t}$ do not necessarily require very large momentum flow
through a $t-$ or $u-$channel propagator, while large top transverse
momenta do; the $\tilde b$ exchange contributions to the exclusive
$t \bar t$ channel are therefore less enhanced at large $m_{t \bar t}$
than at large top $p_T$. As a result of these effects, the bound on
$\lambda^{\prime\prime}_{313}$ is weaker than the limit
(\ref{uni-bound}) even for the smallest $M_{\tilde b}$ considered.  By
comparing the black, red, and blue solid lines we see that the
$t\bar{t}j$ channel dominates the determination of the bound for light
sbottom, while the exclusive $t\bar{t}$ channel dominates for heavy
sbottom.

The SMEFT implementation of our model in terms of $4-$quark operators
yields the the upper bounds
$\lambda^{\prime\prime}_{313}/M_{\tilde{b}} < 2.43$ and $0.89$ when
only the $\Lambda^{-2}$ term and both $\Lambda^{-2} + \Lambda^{-4}$
terms are included in the simulation, respectively, as shown by the
dashed lines. When only the exclusive $t \bar t$ channel is considered
these bounds are much stronger than the corresponding limits derived
in the RPV model. The SMEFT bound derived in linear order is
coincidentally close to the full RPV constraint for the smallest
sbottom mass shown, but has too steep a slope; while the quadratic
SMEFT limit is even stronger than the RPV bound after inclusion of the
$t \bar t j$ channel. The two bounds differ by more than $25\%$ even
at $M_{\tilde b} = 3$ TeV; here the true bound on the coupling lies at
$3.3$, which would lead to a Landau pole at an energy scale of less
than $10$ TeV \cite{Allanach:1999mh}.

\subsection{Transverse momentum of the top pair}

\begin{figure}[htbp]
\centering
\includegraphics[scale=0.355]{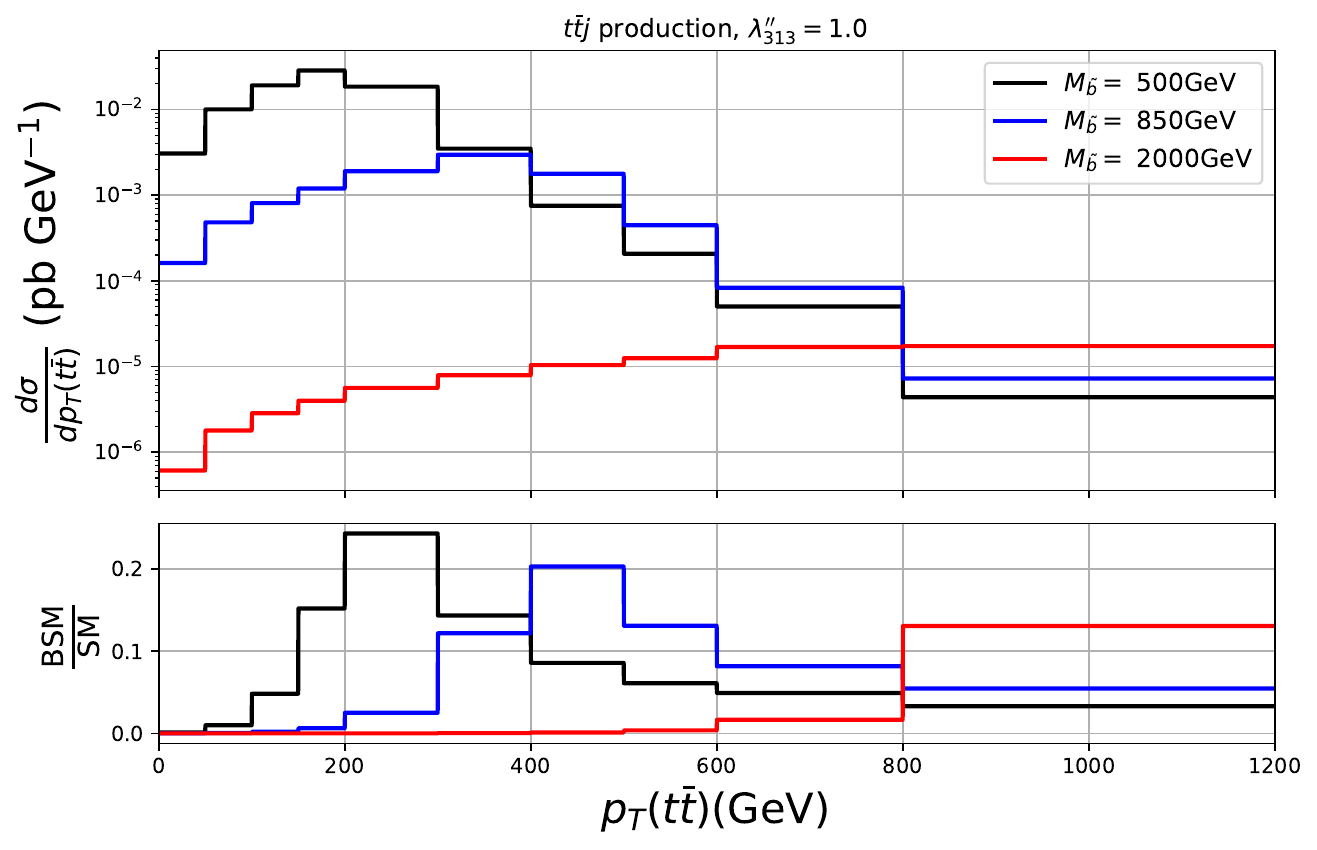}
\hspace{0.1in}
\includegraphics[scale=0.28]{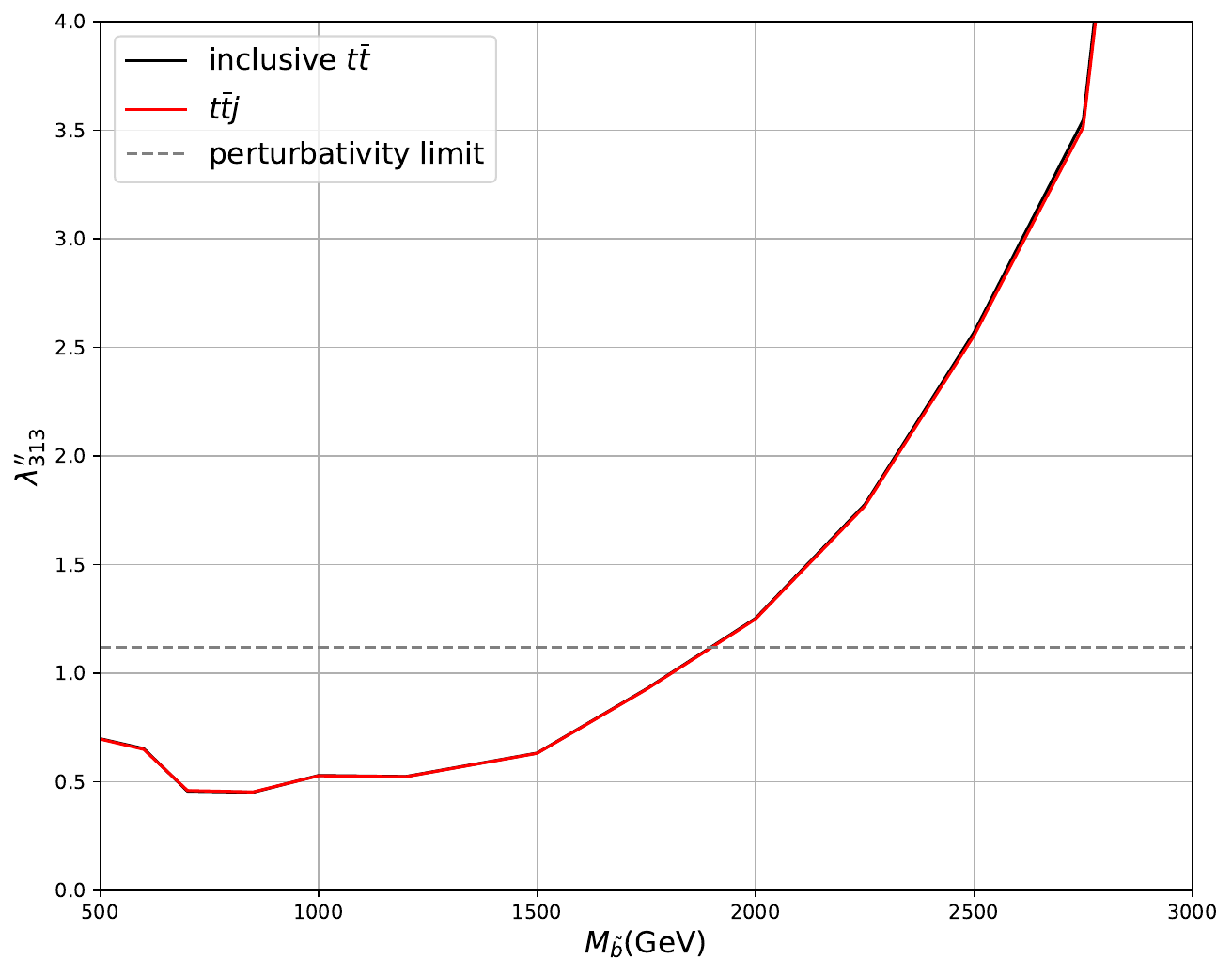}
\caption{Left: parton--level $p_T(t\bar t)$ distribution predicted by
  the RPV model. Right: $95\%$ CL exclusion limits derived on the RPV model
   using the CMS measurement of the $p_T(t\bar t)$ distributions.}
\label{fig:ptt1}
\end{figure}

The production of one or two real sbottom (anti-)squarks gives a
transverse momentum to the $t \bar t$ pair in the final state already
in leading order. The resulting distribution is shown in the left
panel of Fig.~\ref{fig:ptt1}, where we have neglected the (very small)
contribution from the $t \bar t jj$ channel, i.e. only included single
$\tilde b_R$ production. Due to the conservation of transverse
momentum, $p_T({t \bar t})$ is equal to the transverse momentum of the
light quark jet originating from sbottom decay; this distribution
therefore also peaks around $M_{\tilde b}/2$, just like the
contribution from the $t \bar t j$ channel to the $p_T(t_{\rm high})$
distribution does. We also note that increasing the sbottom mass
reduces the value of the cross section at the peak, but increases the
cross section in the last bin. This effect cannot be reproduced by the
SMEFT implementation of the RPV model, where the transverse momentum
of the $t \bar t$ system remains zero in leading order.

The right panel of Fig.~\ref{fig:ptt1} shows the $95\%$ CL exclusion
limit that we derived within the RPV model from the CMS measurement
\cite{CMS:2021vhb} of the parton--level $p_T(t\bar{t})$
distribution. As already noted, in leading order basically only the
$t\bar{t}j$ channel contributes. For $0.55$ TeV $< M_{\tilde b} < 1.24$
TeV the resulting bound on the coupling is slightly weaker than that
we derived from the $p_T(t_{\rm high})$ distribution, see
Fig.~\ref{fig:pthigh_low_bound}. However, it increases less fast for
larger sbottom mass, remaining stronger than the high--scale
perturbativity constraint (\ref{uni-bound}) for $M_{\tilde{b}}< 1.9$
TeV; it thus offers the best sensitivity so far for $1.24$ TeV
$< M_{\tilde b} < 2.1$ TeV. A slightly negative $\Delta\chi^2$ occurs
again for $\lambda^{\prime\prime}_{313} \leq 0.5$ and
$M_{\tilde{b}}\simeq 500$ GeV.

\subsection{Charge asymmetry of the top pair}

Our final observable is the top pair charge asymmetry. Its integrated
value has been measured by ATLAS \cite{ATLAS:2022waa} to be
$0.0068 \pm 0.0015$, which differs from zero by $4.7$ standard
deviations. Following the notation of Ref.~\cite{Czakon:2017lgo}, the
charge asymmetry ($A_C^{t\bar{t}}$) is defined as,
\begin{equation}\label{eq:chargea}
  A_C^{t\bar{t}} = \frac {\sigma_\text{bin}( \Delta |y_{t\bar{t}}|>0)
    - \sigma_\text{bin}(\Delta |y_{t\bar{t}}|<0) }
  {\sigma_\text{bin}(\Delta |y_{t\bar{t}}|>0)
    + \sigma_\text{bin}(\Delta |y_{t\bar{t}}|<0)}
\end{equation}
where $\Delta |y_{t\bar{t}}| = |y_t|-|y_{\bar{t}}|$ is the difference
between the absolute rapidities of top quark and top anti-quark. In
this subsection we analyze the differential measurements of this
quantity as function of $m_{t\bar{t}}$ and of $p_T(t\bar{t})$.

In order to compute the charge asymmetry, we include the SM and BSM
contributions to both the numerator and the denominator in
eq.(\ref{eq:chargea}), where the BSM contribution is again computed
either from from the RPV model or from its SMEFT implementation. The
BSM contributions are obtained from our LO simulations, while the SM
contributions are determined to NNLO QCD with the help of the HighTEA
public tool \cite{Czakon:2023hls}. We use this tool to generate the
cross section (the denominator) with the same PDF set,
renormalization, and factorization scales as those used in the ATLAS
report \cite{ATLAS:2022waa}. Since the parton--level charge asymmetry
from NNLO SM is provided in this report, the SM contribution to the
numerator is computed as the product of the quoted charge asymmetry
and the NNLO cross section computed by us using HighTEA.

\begin{figure}[htbp]
\centering
\includegraphics[scale=0.38]{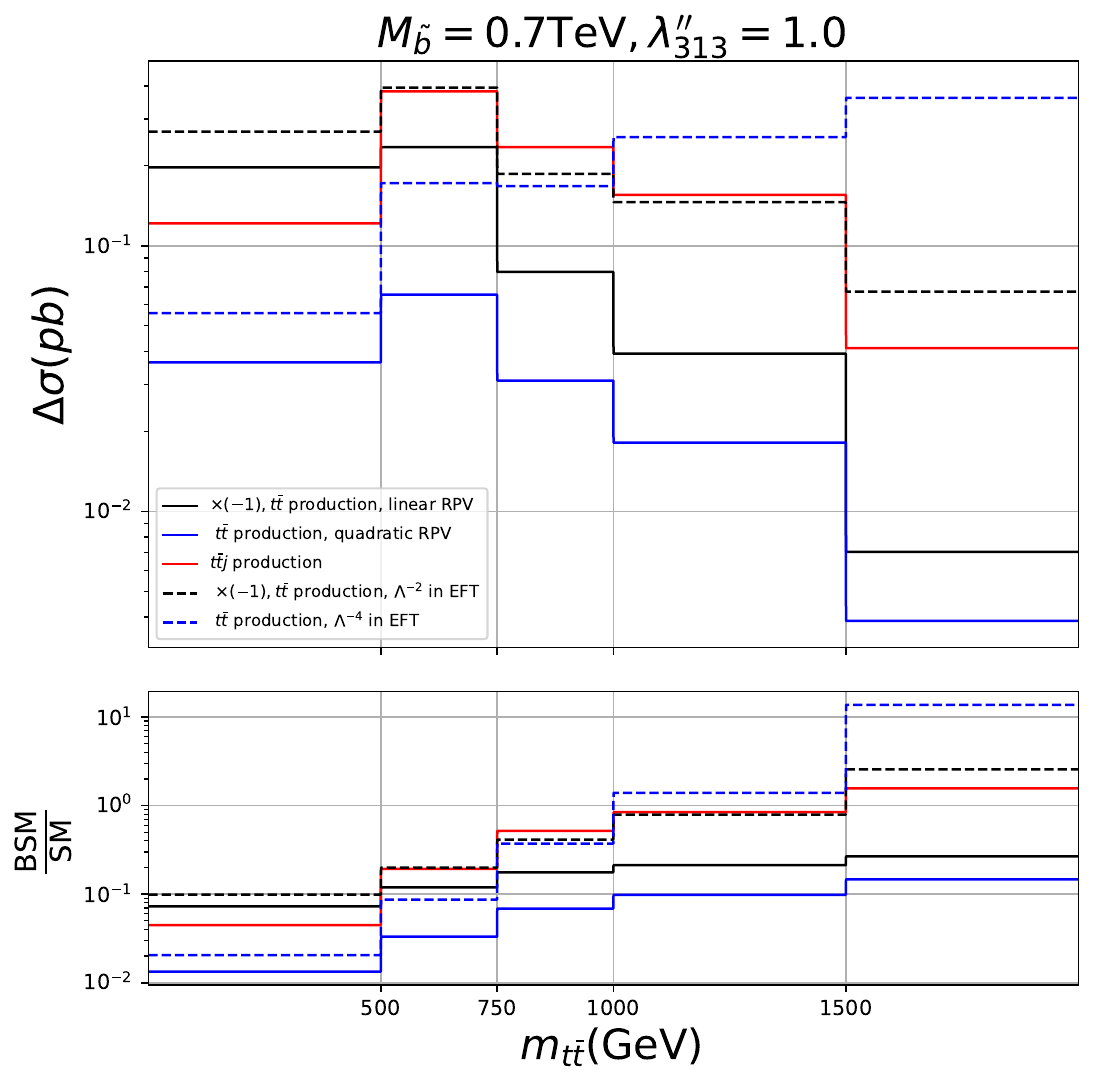}
\hspace{0.1in}
\includegraphics[scale=0.38]{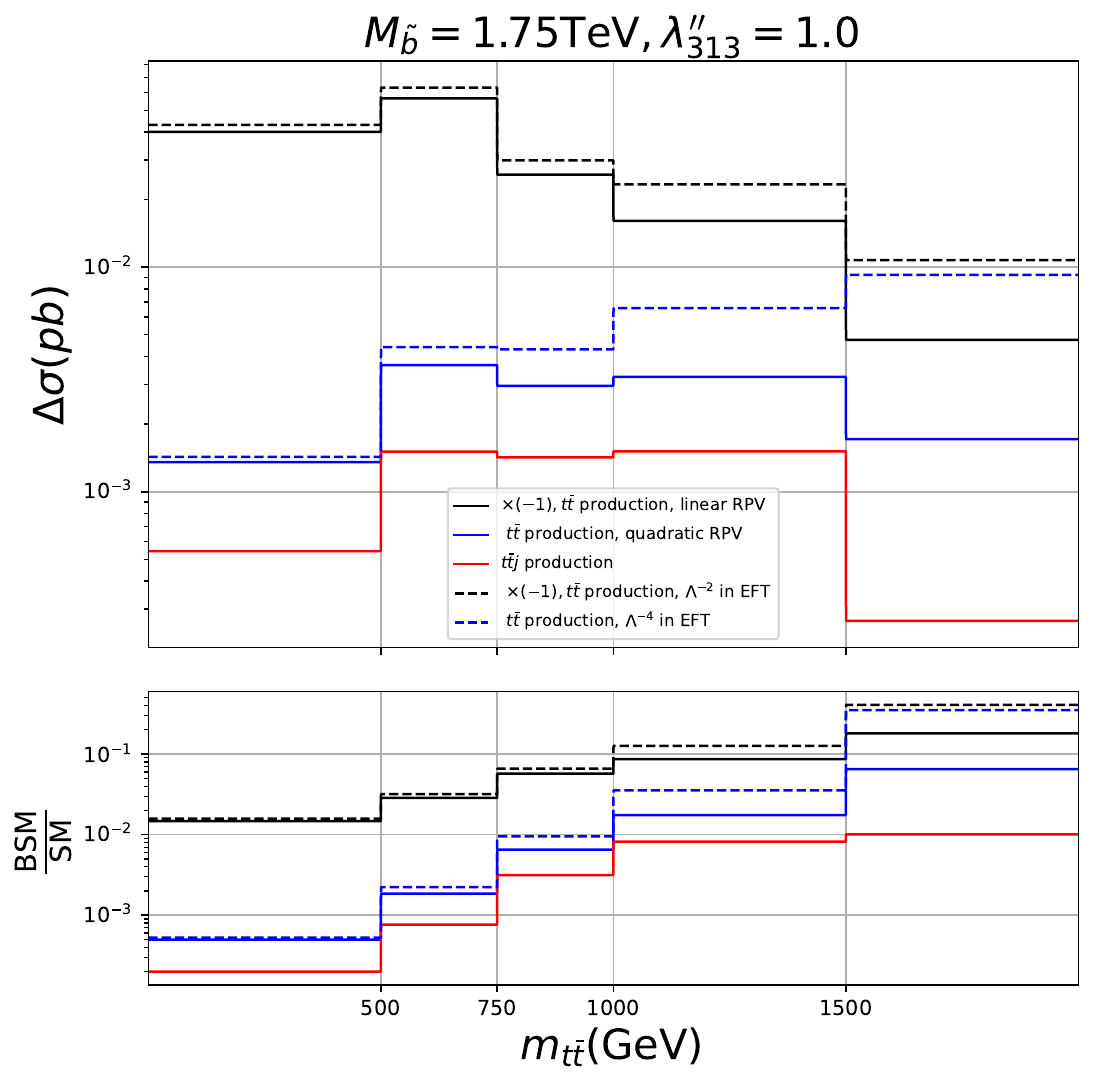}
\caption{The BSM contribution $\Delta\sigma$ to the numerator of
  eq.(\ref{eq:chargea}) as a function of $m_{t\bar{t}}$. The notation is
  as in Fig.~\ref{fig:pthigh_dis}, except that we don't show the
  contribution from the $t \bar t jj$ final state since it vanishes
  in leading order.}
\label{fig:AC_mtt_delta_dis}
\end{figure}

Fig.~\ref{fig:AC_mtt_delta_dis} shows the BSM contribution to the
numerator $\Delta\sigma$ of eq.(\ref{eq:chargea}). These contributions
are non--zero since the four--quark operators given in
eqs.(\ref{wilson}) only include right--handed quarks, i.e. they
contain the chiral projector $P_R$. The resulting $\gamma_5$ terms
give rise to a forward--backward asymmetry in the partonic
center--of--mass frame, which in turn leads to a non--vanishing charge
asymmetry even in leading order. In contrast, in QCD a charge
asymmetry only appears at NLO, and only for top pair production from
quark annihilation; gluon fusion processes do not contribute to
$\Delta \sigma$. To linear order both the RPV model and its SMEFT
implementation predict $\Delta\sigma$ to be negative;
Fig.~\ref{fig:AC_mtt_delta_dis} shows their absolute value. We see
that the contribution from the $t\bar t j$ channel, which only exists
in the RPV model, is quite large for $M_{\tilde b}=0.7$ TeV (left
frames), while it is negligible for $M_{\tilde{b}}=1.75$ TeV. In the
exclusive $t\bar t$ channel the SMEFT implementation again
over--predicts the BSM contribution, particularly in the high bins.
 
\begin{figure}[htbp]
\centering
\includegraphics[scale=0.38]{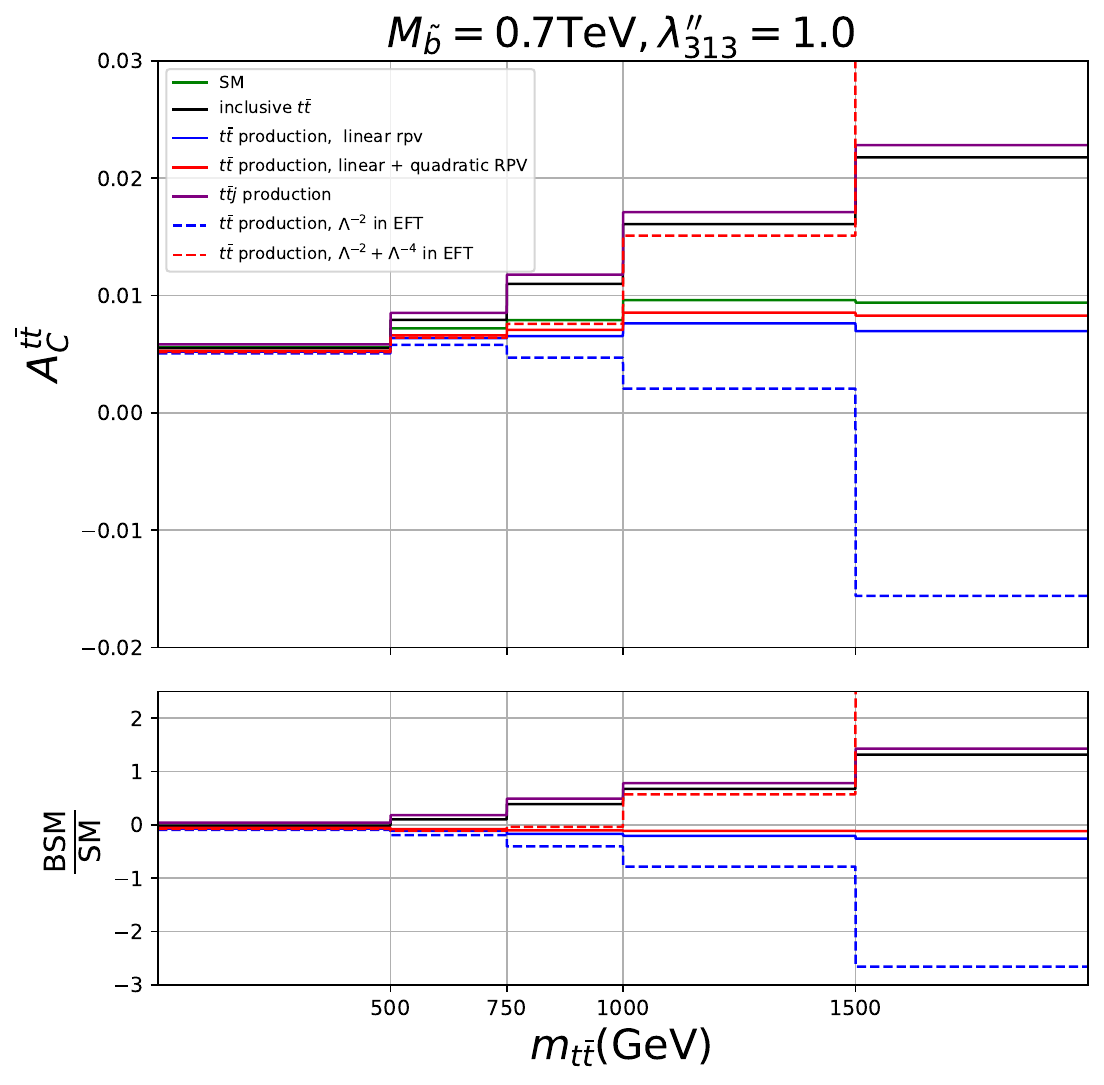}
\hspace{0.1in}
\includegraphics[scale=0.38]{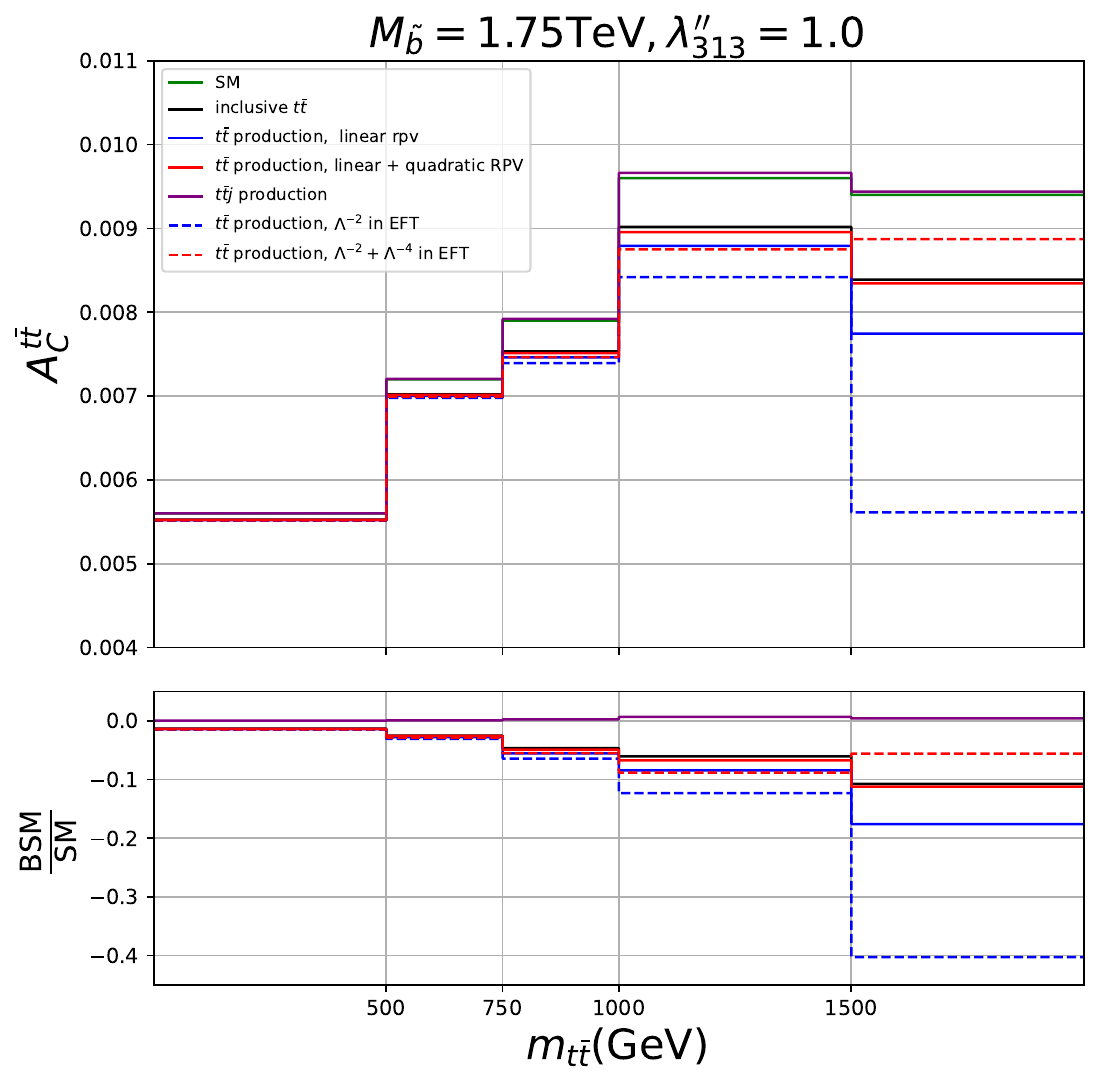}
\caption{Charge asymmetry as function of $m_{t\bar t}$. The green
  histogram shows the SM prediction for $A_C^{t\bar t}$. The other
  histograms depict predictions for $A_{C}^{t\bar t}$ including BSM
  contributions in addition to that from the SM, with
  $\lambda^{\prime\prime}_{313} = 1$. As before, solid and dashed
  histograms show predictions by the RPV model and its SMEFT
  implementation, with blue and red histograms showing predictions for
  the exclusive $t \bar t$ channel to linear and quadratic order,
  respectively. The purple histogram depicts the contribution from the
  $t \bar t j$ channel, and the black histogram shows the complete RPV
  prediction. The lower panels show the difference between BSM and
  predictions normalized to the SM.}
\label{fig:AC_mtt_dis}
\end{figure}

Fig.~\ref{fig:AC_mtt_dis} shows predictions for the charge asymmetry as
function of $m_{t\bar t}$. The SM prediction is shown by the green
histogram; its contribution has been included in all other histograms
as well, which assume $\lambda^{\prime\prime}_{313} = 1$. Including
only linear (i.e. interference) BSM contributions (blue) reduces the
charge asymmetry, but the RPV model predicts it to remain positive
even for $M_{\tilde b} = 0.7$ TeV (left frames). In contrast, the
SMEFT implementation to linear order predicts a negative charge
asymmetry for such a small sbottom mass and large $t \bar t$ invariant
mass. Including the quadratic terms (red) brings the RPV prediction
for the exclusive $t \bar t$ channel quite close to the SM, and leads
to a positive charge asymmetry even in the SMEFT implementation; the
latter becomes quite large (off the scale shown) for
$M_{\tilde b} = 0.7$ TeV.

The $t \bar t j$ channel (violet), which receives LO contributions
only in the RPV model with on--shell $\tilde b_R$ production, is
always positive. Comparison with the black histogram, which shows the
complete prediction in the RPV model, shows that on--shell
$\tilde b_R$ production dominates the charge asymmetry for
$M_{\tilde b} = 0.7$ TeV.  This contribution is suppressed for
$M_{\tilde b} = 1.75$ TeV (right frames). For this combination of
$\tilde b_R$ mass and RPV coupling the total RPV contribution
therefore reduces the charge asymmetry. This is true also in the SMEFT
implementation, which however predicts the difference from the SM
prediction to be too small by nearly a factor of two in the last bin
even for $M_{\tilde b} = 1.75$ TeV.

\begin{table}[htbp]
\centering
\caption{Values of
  $\frac{F(\text{SM+EFT}) - F(\text{SM})} {F(\text{SM}+\text{RPV})
    -F(\text{SM})}$ in the exclusive $t\bar{t}$ channel for two bins
  of $m_{t\bar t}$, two values of the sbottom mass, and two values of
  the RPV coupling $\lambda^{\prime\prime}_{313}$. In the third to
  sixth columns, $F$ is the charge asymmetry, while in the last two
  columns $F$ is $\Delta\sigma$. The upper index $l$ stands for linear
  RPV or SMEFT contributions, while $l+q$ includes the quadratic
  contribution. The values of the coupling are shown in parentheses.
  For example, $A^{l+q}_C(3)$ is the charge asymmetry derived from
  linear + quadratic RPV or SMEFT with
  $\lambda^{\prime\prime}_{313}=3$.}
\label{tab:rpveft_charge}
\begin{tabular}{|c|c|c|c|c|c|c|c|}
\hline
  $M_{\tilde b}$ & $m_{t\bar t}$ & $A^l_C(1)$ & $A^{l+q}_C(1)$ & $A^{l}_C(3)$
  & $A^{l+q}_C(3)$  & $\Delta\sigma^l$ & $\Delta\sigma^{l+q}(1)$   \\
\hline
  \multirow{2}{*}{$0.7$ TeV} & $500-750$ GeV & 1.69 & 1.31& 1.71 & 3.05
                    &1.68 & 1.31  \\
\cline{2-8}
{} & $>1500$ GeV & 10.2 & -72.8& 20.3& 7.30& 9.56 & -92.9  \\
\hline
  \multirow{2}{*}{$1.75$ TeV} & $500-750$ GeV & 1.12 & 1.11& 1.12 & 1.00
                    & 1.12 & 1.11  \\
\cline{2-8}
{}& $>1500$ GeV & 2.29 & 0.5& 2.38& 4.88& 2.27 & 0.5  \\
\hline
\end{tabular}
\end{table}

A more quantitative comparison between the prediction of the full RPV
model and its SMEFT implementation is shown in
Table~\ref{tab:rpveft_charge}, which gives some values of
$\frac{ A_C^{t\bar t}(\text{SM+EFT}) - A_C^{t\bar t}(\text{SM}) } {
  A_C^{t\bar t}(\text{SM}+\text{RPV}) - A_C^{t\bar t}(\text{SM})}$ in
the third to sixth columns and the corresponding values of
$\Delta\sigma(\text{SMEFT})/\Delta\sigma(\text{RPV})$ in the last two
columns. For $M_{\tilde b} = 0.7$ TeV the discrepancy is large even in
the lower invariant mass bin; for $\lambda^{\prime\prime}_{313}=1$
adding the quadratic contributions, which is perfectly reasonable in
the RPV model but less so in its SMEFT implementation, reduces the
discrepancy due to cancellations between the linear and quadratic
contributions. In the higher invariant mass bin the discrepancy
becomes very large. Since in this bin the BSM contribution is sizable
in both the numerator and denominator of the definition
(\ref{eq:chargea}) of the charge asymmetry the ratio of the RPV and
SMEFT predictions for this asymmetry depends on the coupling even to
linear order, whereas the ratio of the predicted $\Delta \sigma$
values (shown in the seventh column) does not.

Increasing the sbottom mass to $1.75$ TeV leads to fairly good
agreement between the predictions in the lower invariant mass bin;
however, in the bin with large invariant mass, where the deviation
from the SM prediction is much more prominent as shown in the lower
frames of Fig.~\ref{fig:AC_mtt_dis}, the two predictions still differ
by a factor of $2$ or more.

\begin{figure}[h]
\centering
\includegraphics[scale=0.6]{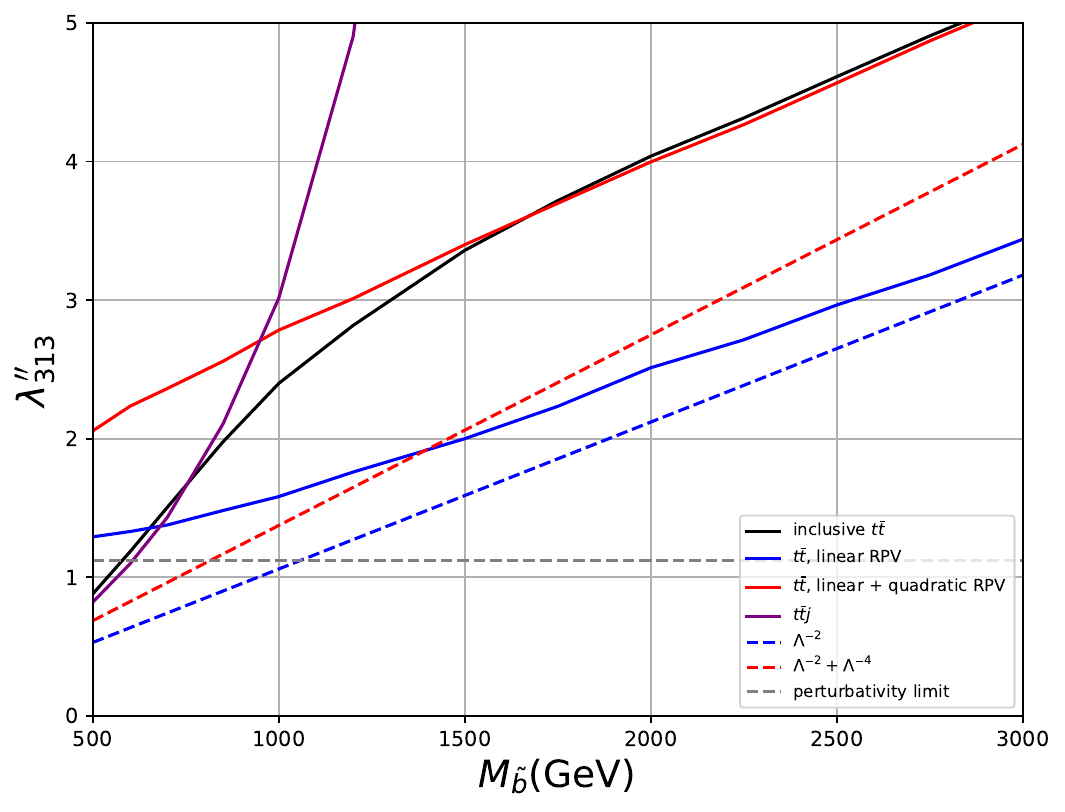}
\caption{Similar to Fig.~\ref{fig:pthigh_low_bound}, but for exclusion
  limits derived from the ATLAS measurement \cite{ATLAS:2022waa} of
  the charge asymmetry as a function of $m_{t\bar t}$. The SMEFT
  implementation of our RPV model leads to the upper bounds
  $\lambda^{\prime\prime}_{313}/M_{\tilde{b}}<1.06$ ($1.37$) in linear
  (linear plus quadratic) order, respectively, as shown by the dashed
  lines.}
\label{fig:AC_mtt_bound}
\end{figure}

Fig.~\ref{fig:AC_mtt_bound} presents $95\%$ CL exclusion limits
obtained from the ATLAS measurement \cite{ATLAS:2022waa} of the
parton--level charge asymmetry as a function of $m_{t\bar t}$. The
complete RPV model leads to a rapidly weakening bound as the sbottom
mass in increased. The $t\bar{t}j$ channel dominates the determination
of the bound for $M_{\tilde b} \leq 750$ GeV, while the exclusive
$t\bar{t}$ channel dominates for heavy sbottom. Overall the bound is
quite weak, excluding couplings below the bound (\ref{uni-bound}) from
demanding perturbative unitarity up to very high energy scales only
for $M_{\tilde b} < 570$ GeV. Moreover, the SMEFT implementation again
leads to much too strong constraints for the entire range of sbottom
mass shown. In contrast to the bounds derived from the $p_T$ spectrum
of top (anti--)quarks or from the $t \bar t$ invariant mass spectrum,
including quadratic SMEFT contributions weakens the constraint, since
they don't suffice to flip the sign of the linear (interference)
contribution even in the highest bins.

\begin{figure}[h]
\centering
\includegraphics[scale=0.35]{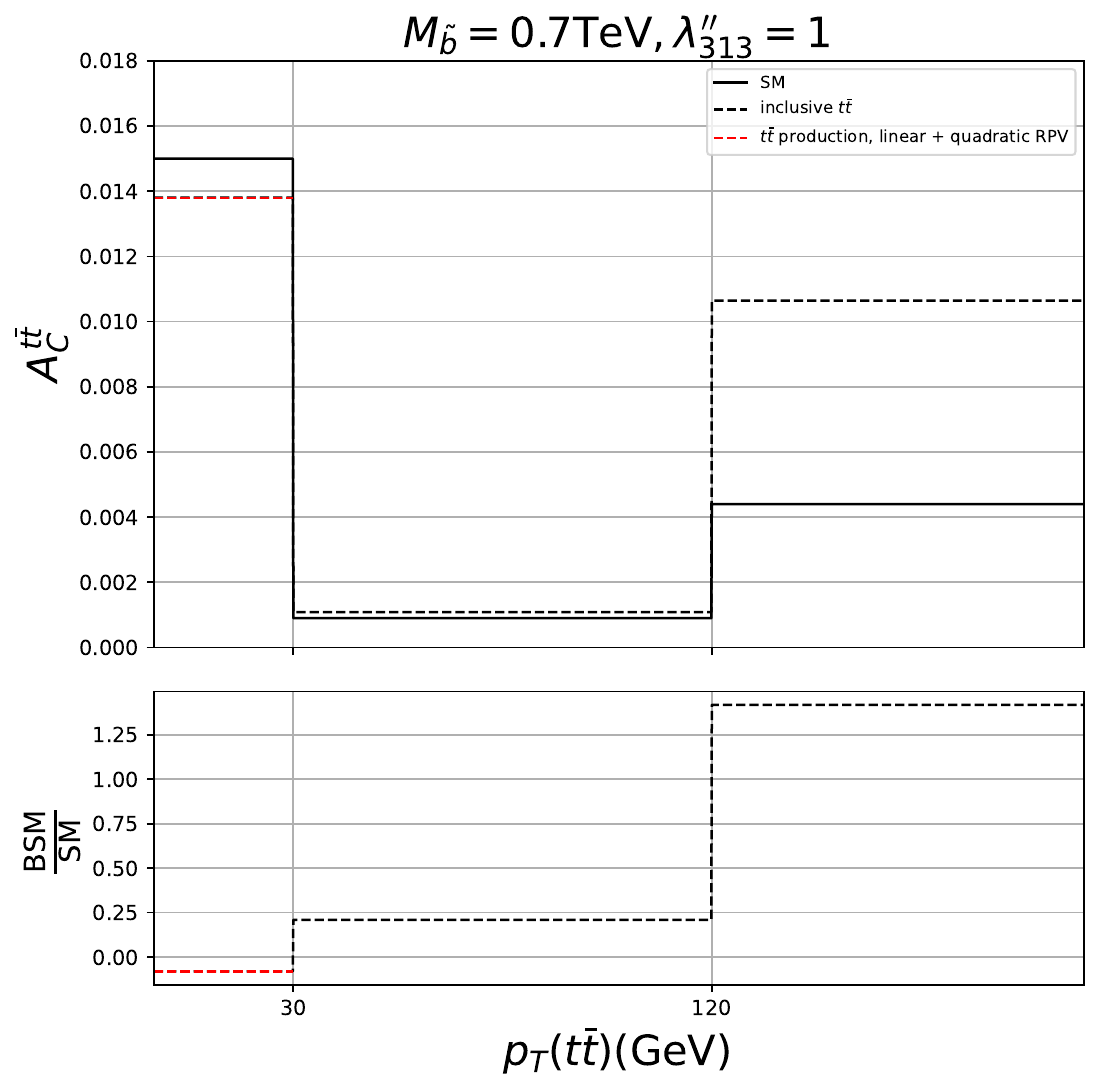}
\hspace{0.1in}
\includegraphics[scale=0.45]{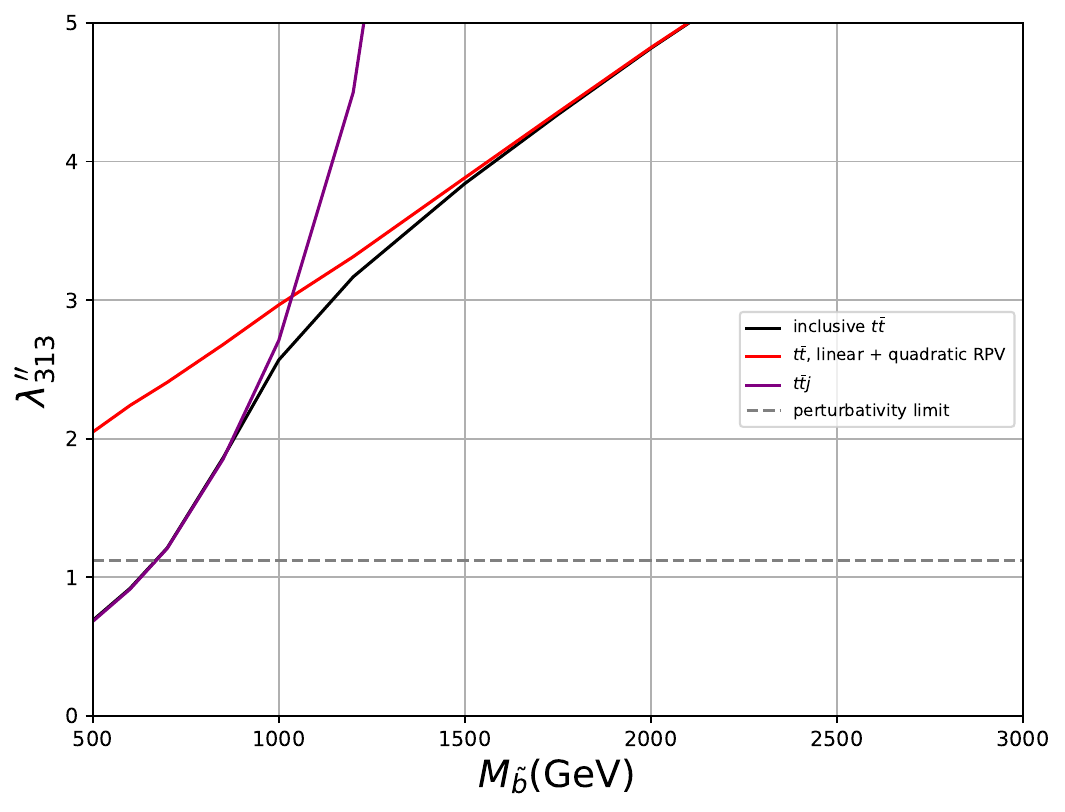}
\caption{Left: the charge asymmetry defined in eq.(\ref{eq:chargea})
  as function of $p_T(t\bar t)$, as predicted in the SM (solid) and in
  the RPV theory (dashed). The dashed red histogram is for the
  exclusive $t \bar t$ channel, which to leading order only
  contributes in the first bin; in this bin it overlaps with the
  dashed black histogram which shows the RPV prediction for inclusive
  $t \bar t$ production. Right: the $95\%$ CL exclusion limit derived
  in the framework of the RPV model from the ATLAS measurement
  \cite{ATLAS:2022waa} of the the charge asymmetry as a function of
  $p_T{(t\bar t)}$.}
\label{fig:AC_ptt}
\end{figure}

The left panel of Fig.~\ref{fig:AC_ptt} presents the predicted charge
asymmetry as a function of $p_T(t\bar t)$ for $M_{\tilde b}= 0.7$ TeV
and $\lambda^{\prime\prime}_{313} = 1$.  Sbottom exchange to the
exclusive $t \bar t$ channel contributes to leading order only at
$p_T(t\bar t)=0$, as shown by the red dashed histogram. The deviations
of $A_{C}^{t\bar t}$ in the second and third bins are thus entirely
due to the $t\bar{t}j$ channel. Since this contribution is peaked at
$p_T(t \bar t) \simeq M_{\tilde b}/2$ it is most prominent in the
highest bin.

Recall that the QCD prediction is at NNLO, and therefore extends to
non--vanishing $p_T(t \bar t)$. Since gluons emit more initial state
radiation than quarks do, the gluon fusion channel, which does not
contribute to the numerator of the charge asymmetry, is less important
in the lowest bin, but completely dominates the second bin; this
explains the steep decline of the SM prediction between these two
bins. At very large values of $p_T(t \bar t)$ the contribution from
$q \bar q$ annihilation becomes somewhat more important again since
the valence quark distributions are harder than the gluon distribution
in the proton; as a result the SM prediction increases again in the
third, and highest, bin.

The right panel of Fig.~\ref{fig:AC_ptt} depicts $95\%$ CL exclusion
limits obtained from the ATLAS measurement \cite{ATLAS:2022waa} of the
parton--level charge asymmetry as a function of $p_T(t\bar t)$. The
complete RPV bound (black) is essentially determined by the
$t \bar t j$ channel (purple) for $M_{\tilde b} < 1$ TeV. However, for
$M_{\tilde b} \leq 0.9$ TeV the bound on the RPV coupling is only
slightly stronger than that derived from the distribution of the
charge asymmetry as function of $m_{t \bar t}$, see
Fig.~\ref{fig:AC_mtt_bound}, and even weaker for larger sbottom mass.
The bound falls below the perturbative unitarity limit
(\ref{uni-bound}) only for $M_{\tilde b}< 670$ GeV.

\subsection{Summary of exclusion limits in the RPV model and its SMEFT
  implementation}

\begin{figure}[h]
\centering
\includegraphics[scale=0.5]{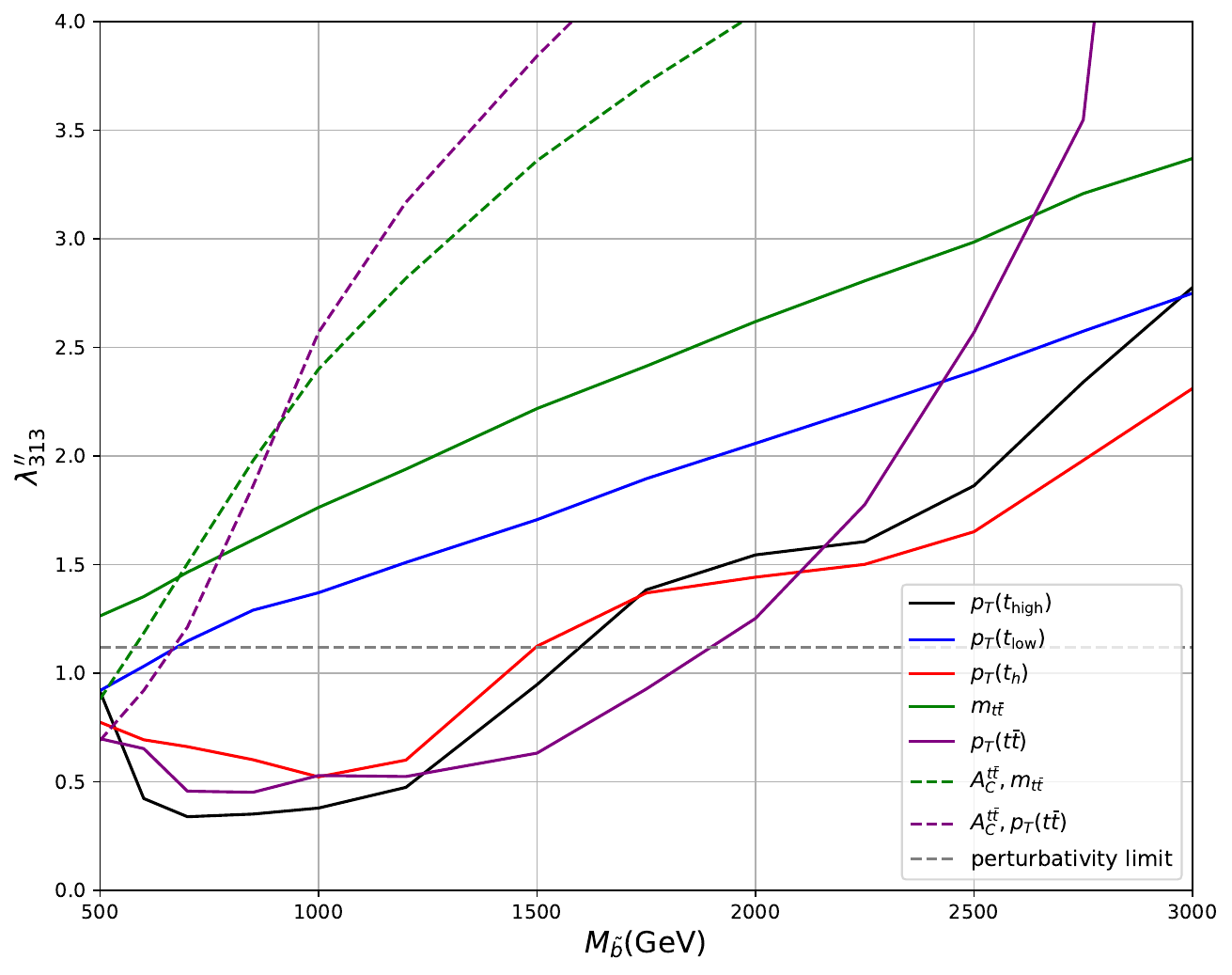}
\caption{Summary of $95\%$ CL exclusion limits on the RPV coupling as
  function of the sbottom mass. The solid curves have been derived
  from CMS measurements \cite{CMS:2021vhb} of the inclusive $t \bar t$
  production cross section differential in some transverse momentum or
  in the $t \bar t$ invariant mass, while the dashed green and purple
  curves have been derived from ATLAS measurements
  \cite{ATLAS:2022waa} of the charge asymmetry in inclusive $t \bar t$
  production. The dashed gray line shows the bound (\ref{uni-bound})
  from demanding perturbative unitarity up to very high energy scales;
  for $\lambda^{\prime\prime}_{313} > 3$ perturbation theory becomes
  questionable even at the scales probed at the LHC.}
\label{fig:bound_summary}
\end{figure}

Fig.~\ref{fig:bound_summary} collects the limits on the RPV coupling
derived in the full RPV model as discussed in the previous
subsections. We see that the strongest upper bound on the coupling
comes from either the $p_T(t_{\rm high})$ distribution (solid black),
the $p_T(t_h)$ distribution (solid red) or the $p_T(t\bar t)$
distribution (solid purple), depending on the value of $M_{\tilde b}$.
Since these three bounds have been derived from the same data set
\cite{CMS:2021vhb} their statistical combination is not
straightforward; lacking such a combination one can simply take the
strongest of these three limits as final bound derived from the CMS
data.  The resulting upper bound on $\lambda^{\prime\prime}_{313}$ is
stronger than the constraint (\ref{uni-bound}) from high--scale
unitarity for $M_{\tilde b} < 1.9$ TeV.

In Fig.~\ref{fig:bound_summary} we extend the constraints to
$M_{\tilde b} = 3$ TeV and allow values of the coupling
$\lambda^{\prime\prime}_{313} \leq 4$.  It should be noted that the
Breit--Wigner propagator, with constant (energy independent) width, is
used for the unstable sbottom in the $t\bar{t}j$ and $t\bar{t}jj$
channels. This becomes a rather poor approximation for
$\lambda^{\prime\prime}_{313}\gtrsim 3$, as mentioned in
Sec.~\ref{sec2}.  One reason is that the width originates from the
imaginary part of the $\tilde b_R$ two--point function, which depends
on the square of the four--momentum $q^2$ flowing through it; if the
width is a sizable fraction of $M_{\tilde b}$ it will remain important
over an extended range of $q^2$, where the imaginary part varies
considerably. For $q^2 \sim M^2_{\tilde b}$ these diagrams are
therefore specially sensitive to loop corrections.

In the $\tilde b_R$ exchange contribution to exclusive $t \bar t$
production, described by the first diagram in
Fig.~\ref{fig:feyn-diagram}, the exchanged momentum is space--like,
hence the corresponding two--point function has no imaginary
part. However, the real part of the two--point function will modify
the $t-$channel $\tilde b_R$ propagator at one--loop
order.\footnote{These are the only
  ${\cal O}\left[\left( \lambda^{\prime\prime}_{313}\right)^4 g_S^2
  \right] $ corrections to $\sigma(d \bar d \rightarrow t \bar t) $;
  at two--loop order additional
  ${\cal O}\left[ \left(\lambda^{\prime\prime}_{313}\right)^6 g_S^2
  \right]$ corrections appear, e.g. from double box diagrams. There
  are also one--loop
  ${\cal O}\left[\left( \lambda^{\prime\prime}_{313}\right)^4 g_S^2
  \right]$ box diagram corrections to
  $\sigma(d \bar d \rightarrow d \bar d)$. However, in the absence of
  efficient ($d-$)quark tagging these would contribute to inclusive
  jet production, which is dominated by $gg \rightarrow gg$
  scattering; the impact of our RPV coupling on inclusive jet
  production is therefore much less than that on $t \bar t$
  production.} Recall also that the $t \bar t j$ channel affects the
bound on $\lambda^{\prime\prime}_{313}$ even for
$M_{\tilde b} \simeq 3$ TeV. Therefore the bounds shown in
Fig.~\ref{fig:bound_summary} may not be very reliable for
$\lambda^{\prime\prime}_{313}>3$. Since our LO calculation almost
certainly cannot be trusted for $\lambda^{\prime\prime}_{313}>4$ we do
not extend our bounds to such large couplings.

Recall that the three observables giving the best bounds on our RPV
coupling are not described well by the SMEFT implementation of the RPV
model. In leading order $p_T(t\bar t)$ gets contributions only from
the $t \bar t j$ and $t \bar t jj$ channels, which do not exist in the
SMEFT implementation. The $t \bar t j$ channel also makes significant
contributions to $p_T(t_{\rm high})$ and $p_T(t_h)$ even at
$M_{\tilde b} = 3$ TeV. The discrepancy between the RPV model and its
SMEFT implementation was smaller in the $p_{T}(t_{\rm low})$
distribution, which however yields a much weaker constraint on
$\lambda^{\prime\prime}_{313}$.

\begin{table}[htbp]
\centering
\caption{$95\%$ CL bounds on the Wilson coefficients of the two $d=6$
  SMEFT operators generated at tree--level from the RPV model, in
  units of $\text{TeV}^{-2}$. The second and third column are taken
  from ref.\cite{Ethier:2021bye}; ``Individual'' means that only a
  single Wilson coefficient is allowed to be nonzero, whereas
  ``Marginalized'' refers to the results of a fit where up to $50$
  coefficients are allowed to float. The next three columns show
  results derived by us, as described earlier in this section. The
  last column shows individual results from the ATLAS collaboration
  \cite{ATLAS:2022waa}, using their measurement of the charge
  asymmetry.}
\label{tab:smeft_fit}
\scalebox{0.95}{
\begin{tabular}{|c|c|c|c|c|c|c|}
\hline
  Operator &Individual  & Marginalized & $p_T(t_h)$ & $m_{t\bar{t}}$ &
$A_{C}^{t\bar{t}}(m_{t\bar{t}})$ &$A_{C}^{t\bar{t}}(m_{t\bar{t}})$,
 ATLAS \\
\hline
\multicolumn{7}{|c|}{$\Lambda^{-2}$}\\
\hline
$\mathcal{O}_{t d}^{(1)}$ & [-9.504,-0.086] &[-27.673,11.356]& &&&[-1.94,1.00]\\
\hline
  $\mathcal{O}_{t d}^{(8)}$ & [-1.458,1.365]  &[-5.494,25.358] &-1.300 &-5.905
          &-1.124&[-0.45,2.13]\\
\hline
\multicolumn{7}{|c|}{$\Lambda^{-2}+\Lambda^{-4}$}\\
\hline
  $\mathcal{O}_{t d}^{(1)}$ & [-0.449,0.371] &[-0.474,0.347] & 0.188 & 0.264 &
       0.626 & [-0.60,0.84]\\
\hline
  $\mathcal{O}_{t d}^{(8)}$ & [-1.308,0.638] & [-1.329,0.643] & -0.563 & -0.792
        &-1.877&[-1.62,1.21]\\
\hline
\end{tabular}}
\end{table}

For completeness we nevertheless collect in Table~\ref{tab:smeft_fit}
the bounds on the Wilson coefficients of the two $d=6$ SMEFT operators
that are generated at tree--level by our RPV model. The second and
third columns show fit results from ref.\cite{Ethier:2021bye}, which
predates the publication of the CMS and ATLAS data we used for our
analysis. The next three columns show the constraints we derived from
the measured $p_T(t_h),\ m_{t \bar t}$ and
$A_C^{t \bar t}(m_{t \bar t})$ distributions, respectively, and the
last column shows the constraints derived by ATLAS
\cite{ATLAS:2022waa} from the latter distribution. Note that in our
analysis we always assume $C^1_{td} = -C^8_{td}/3 > 0$, see
eqs.(\ref{wilson1}), whereas in refs.\cite{Ethier:2021bye} and
\cite{ATLAS:2022waa} these two coefficients are assumed to be
independent and are allowed to have either sign. Moreover, we only
include $d$ (anti--)quarks in the initial state, whereas
refs.\cite{Ethier:2021bye} and \cite{ATLAS:2022waa} assume equal
couplings for $d$ and $s$ quarks; however, since there are no strange
valence quarks in the proton the additional contribution from
$s \bar s$ initial states is quite small.

A final difference is that we only consider LO QCD matrix elements
when computing the interference with SMEFT operators (or with the full
matrix element predicted by the RPV model), whereas
refs.\cite{Ethier:2021bye} and \cite{ATLAS:2022waa} also include
electroweak contributions to the SM amplitudes. This explains why
these references obtain a bound on $C^1_{td}$ already a linear order;
recall that ${\cal O}^{(1)}_{td}$ does not interfere with the LO QCD
contribution to $d \bar d \rightarrow t \bar t$, due to the different
color structure. However, these bounds are not very strong.

We see that the bound on $C^8_{td}$ that we derive to linear order
from either the $p_T(t_h)$ or the $A_C^{t \bar t}(m_{t \bar t})$
distributions is comparable to the corresponding individual constraint
from ref.\cite{Ethier:2021bye}, remembering that we only allow
negative values for this coefficient. This indicates that including
these observables into a global SMEFT fit might have some impact even
on the individual fit; the impact would be much greater if these
measurements break some degeneracy, which must be responsible for the
greatly weakened bounds in the marginalized fit. We also note that the
ATLAS constraint on this coefficient is considerably stronger on the
negative side than what we find. This is probably because they define
their allowed interval relative to the best--fit value, which is
positive in this case. We define our $\Delta \chi^2$ relative to the
SM prediction (which is also the best--fit value after imposing our
constraint $C^8_{td} \leq 0$).

We also see that including the $\Lambda^{-4}$ contributions,
i.e. performing a quadratic fit, greatly changes the constraints. In
this case the constraint on $C^8_{td}$ we derive from the $p_T(t_h)$
or $m_{t \bar t}$ distribution is considerably stronger even than the
individual fits in the literature; since
$\left| C^1_{td} \right|^2 = \left| C^8_{td} \right|^2/9$ the fact
that we include two operators probably only plays a comparatively
minor role here. The inclusion of the differential $t \bar t$
distributions measured by the CMS collaboration should therefore have
a sizable impact on the quadratic fit even if the methodology of
ref.\cite{Ethier:2021bye} is adopted.

However, the large difference between the results of the linear and
quadratic fits also shows that these bounds are not reliable even
within the framework of the SMEFT. As we emphasized previously, to
${\cal O}(\Lambda^{-4})$ one should also include the interference of
$d=8$ SMEFT operators with the SM contribution, at least in the
marginalized fit. Some of these contributions can surely be negative,
whereas the square of $d=6$ operators can obviously only increase
cross sections. More fundamentally, finding large differences in
${\cal O}(\Lambda^{-2})$ and ${\cal O}(\Lambda^{-4})$ fits shows that
the expansion in inverse powers of $\Lambda$ does not converge when
applied to current LHC data, casting doubt on the principle by which
the SMEFT is constructed.

\section{Conclusions}
\label{sec5}

In this paper we tested how well a simplified supersymmetric model
with $R-$parity violation can be described by $d=6$ $4-$quark
operators contained in the SMEFT. Specifically, we considered
contributions from $\tilde b_R$ exchange to inclusive top pair
production. This analysis is facilitated by the fact that CMS
\cite{CMS:2021vhb} and ATLAS \cite{ATLAS:2022waa} show various
measured distributions at the parton level, which we can directly
compare to parton--level simulations.\footnote{Here we are assuming
  that BSM effects do not significantly alter the reconstruction of
  parton--level observables from the actual experimental
  measurements.} We assume that the $\tilde b_R$ is considerably
lighter than all other supersymmetric particles, so that their
production can be ignored; and that the only non--zero new coupling is
$\lambda_{313}^{\prime\prime}$. These assumptions are not particularly
``natural'' from the supersymmetric model building point of view, but
they are ``SMEFT friendly'', since the latter cannot be expected to
correctly model the production of (for example) on--shell
gluinos.\footnote{The cross section for gluino pair production or
  associate gluino plus first generation squark production remains
  significant for masses up to $2.5$ TeV at least \cite{Drees:2004jm,
    Baer:2006rs}. When considering $\tilde b_R$ masses down to $0.5$
  TeV we therefore implicitly assume a sizable hierarchy between the
  masses of strongly interacting superparticles, which tends to be
  destroyed by renormalization group running \cite{Drees:2004jm,
    Baer:2006rs}.} We saw in Sec.~\ref{sec4} that direct searches do
not seem to constrain this scenario strongly. We therefore consider
sbottom masses from $0.5$ TeV upwards.

The coefficients of the relevant $4-$ quark operators
${\cal O}^{(1)}_{td}$ and ${\cal O}^{(8)}_{td}$ can easily be obtained
by integrating out $\tilde b_R$ at tree level, see
eqs.(\ref{wilson1}). Clearly this SMEFT implementation of the RPV model
can only work if the production of on--shell $\tilde b_R$ squark or
antisquarks does not contribute significantly. However, we found that
the distribution of the top + jet invariant mass ($m_{tj}$) in the
$t\bar{t}j$ channel exhibits a significant resonance peak even for a
sbottom mass as high as 3 TeV, see Fig.~\ref{fig:resonance_dis}. This
channel not only completely dominates the $p_T(t \bar t)$
distribution, see Fig.~\ref{fig:ptt1}, where exclusive $t \bar t$
production does not contribute at leading order; due to the Jacobian
peak at $M_{\tilde b}/2$ it also remains very significant in the $p_T$
distributions of single top quarks, see Figs.~\ref{fig:pthigh_dis} and
\ref{fig:ptbound}. These happen to be the distributions that give the
strongest constraints on the RPV model, see
Fig.~\ref{fig:bound_summary}. It is therefore clear that these
constraints can {\em not} be reproduced by the SMEFT implementation of
the RPV model, as shown explicitly in Fig.~\ref{fig:ptbound}.

Moreover, even if we discard the $t \bar t j$ channel by focusing on
the exclusive (parton--level) $t \bar t$ channel the predictions by
the SMEFT implementation differ considerably from those of the RPV
model for all sbottom masses of current interest, in particular in the
bins with high transverse momentum or high invariant mass which are
most sensitive to these BSM effects; see Tables~\ref{rpveft},
\ref{rpveft2} and \ref{tab:rpveft_charge}. Here the SMEFT implementation
overestimates the size of the BSM contributions, since it ignores the
momentum flow through the exchanged $\tilde b_R$ squark, i.e. replaces
the propagator $1/(q^2 - M^2_{\tilde b})$ by $-1/M^2_{\tilde b}$; this
is a bad approximation once $|q^2|$ becomes of order $M^2_{\tilde
  b}$.

The quality of the SMEFT approximation therefore evidently depends on
the mass of the sbottom; the SMEFT implementation will certainly
become reliable at some value of $M_{\tilde b}$. However, very heavy
sbottom squarks lead to measurable effects only for very large RPV
coupling, where perturbation theory breaks down. In fact, demanding
the coupling to remain perturbative all the way up to the scale of
Grand Unification leads to the constraint (\ref{uni-bound}). If this
bound is implemented the data we analyzed only impose new limits for
$M_{\tilde b} < 1.9$ TeV, where the SMEFT implementation fails badly.
For $M_{\tilde b} = 3$ TeV these data can only exclude scenarios with
coupling $\lambda_{313}^{\prime\prime} > 2.3$, where the reliability of
perturbation theory at the LHC scale is already somewhat questionable;
we just saw that the SMEFT implementation still does not work at this
sbottom mass.

In addition to the transverse momentum and invariant mass
distributions measured by CMS, we also analyzed charge asymmetry
distributions measured by ATLAS \cite{ATLAS:2022waa}. Here, too, the
SMEFT implementation fails, see Fig.~\ref{fig:AC_mtt_dis}; the bounds
derived from it are much stronger than the ones derived in the RPV
model for all $M_{\tilde b} \leq 3$ TeV, see Fig.~\ref{fig:AC_mtt_bound}.
The bounds derived in the RPV model are considerably weaker than the
ones derived from the CMS data.

Even for $M_{\tilde b} < 3$ TeV the SMEFT implementation does not fail
everywhere. By focusing on the exclusive $t \bar t$ channel and
removing events with high top transverse momentum and/or high
$t \bar t$ invariant mass one should be able to define a region of
phase space where the absolute value of the squared momentum flowing
through the $\tilde b_R$ propagator is much smaller than
$M^2_{\tilde b}$. However, this procedure would remove those events
which are {\em most} sensitive to the BSM effects we studied here;
surely this is not the appropriate algorithm for deriving bounds on
extensions of the Standard Model.

We therefore conclude that there is {\em no} region of parameter space
of our RPV model where it simultaneously leads to effects that are
measurable at the LHC and can be described well by a SMEFT
implementation. Once mild constraints on perturbativity are imposed
the former is true only for $M_{\tilde b} \leq 3$ TeV, where the
predictions of the RPV model differ significantly from those by its
SMEFT implementation, especially in the tails of distributions which
are most sensitive to BSM effects.

This is true even though we attempted to construct a SMEFT friendly
scenario. In particular, there is essentially no $2 \rightarrow 1$
$\tilde b_R$ resonance production in our model, which could require a
top quark distribution in the proton; this is to be contrasted with
models containing new gauge bosons, or many diquark models, where such
resonance production is the dominant discovery channel, which can of
course not be described by a SMEFT implementation. On the other hand,
$\tilde b_R$ production does contribute at tree--level to the
production of final states containing only SM particles, in contrast
to $R-$parity conserving supersymmetry whose dominant LHC signals
certainly cannot be modeled by the SMEFT. The fact that the SMEFT
cannot even describe our favorable scenario indicates that the SMEFT
is model independent only in the sense that it does not reproduce the
LHC signals of {\em any} perturbative model.

In fact, at least for the operators we considered the situation is
worse than this. We saw in Table~\ref{tab:smeft_fit} that the
constraints on their Wilson coefficients very strongly depend on
whether or not the contributions proportional to the squares of these
operators are included. A ${\cal O}(\Lambda^{-2})$ fit thus yields
very different results from a fit that includes those
${\cal O}(\Lambda^{-4})$ terms that can be computed from the same
Wilson coefficients, showing that the expansion in inverse powers of
$\Lambda$ does not converge. This invalidates the very ansatz on which
the SMEFT is constructed.

\acknowledgments

We are grateful to CMS collaboration for providing the data on NNLO SM
predictions, including the differential cross sections and covariance
matrices, and to ATLAS collaboration for updating their covariance
matrices.

\bibliographystyle{jhep}
\bibliography{ref}
\end{document}